\documentclass[preprint,authoryear]{elsarticle}

\RequirePackage{amsthm,amsmath,amsfonts,amssymb}
\RequirePackage[authoryear]{natbib}
\usepackage{pgffor} 
\usepackage[font=small]{caption}
\usepackage{subcaption}
\usepackage{bbm}
\usepackage{xcolor}
\usepackage{geometry}
\usepackage{float} 
\usepackage[tableposition=top]{caption}
\setcitestyle{square, numbers,sort}
\usepackage[colorlinks = true,
            linkcolor = blue,
            urlcolor  = blue,
            citecolor = blue,
            anchorcolor = blue]{hyperref}
\usepackage{natbib}

\usepackage{soul}
\setcitestyle{authoryear,open={(},close={)}}
\usepackage{array}
\newcolumntype{Y}{>{\raggedright\arraybackslash}p{2.3cm}}
\newcolumntype{Z}{>{\raggedleft\arraybackslash}p{2.8cm}}

\usepackage{multirow}

\usepackage{booktabs}
\usepackage{xcolor} 

\journal{International Journal of Forecasting}

\begin{document}

\begin{frontmatter}

\title{Restoring the Forecasting Power of Google Trends \\
with Statistical Preprocessing}

\author[1]{Candice Djorno}

\author[2]{Mauricio Santillana}

\author[1]{Shihao Yang}


\affiliation[1]{organization={H. Milton Stewart School of Industrial and Systems Engineering, Georgia Institute of Technology}, city={Atlanta}, state={GA}, country={USA}}
\affiliation[2]{organization={Department of Physics, Northeastern University}, city={Boston}, state={MA}, country={USA}}


\begin{abstract} 
Google Trends reports how frequently specific queries are searched on Google over time. It is widely used in research and industry to gain early insights into public interest. However, its data generation mechanism introduces missing values, sampling variability, noise, and trends. These issues arise from privacy thresholds mapping low search volumes to zeros, daily sampling variations causing discrepancies across historical downloads, and algorithm updates altering volume magnitudes over time. Data quality has recently deteriorated, with more zeros and noise, even for previously stable queries. We propose a comprehensive statistical methodology to preprocess Google Trends search information using hierarchical clustering, smoothing splines, and detrending. We validate our approach by forecasting U.S. influenza hospitalizations up to three weeks ahead with several statistical and machine learning models. Compared to omitting exogenous variables, our results show that preprocessed signals enhance forecast accuracy, while raw Google Trends data often degrades performance in statistical models.
\end{abstract}

\begin{keyword}
Google Trends \sep Preprocessing \sep Clustering \sep Denoising \sep Detrending \sep Forecasting
\end{keyword}

\end{frontmatter}

\section{Introduction}
\label{sec:introduction}

The digital footprint generated by billions of daily online searches offers a wealth of data, capturing public interests and behaviors across a broad range of topics. As the world’s leading search engine, Google plays a central role in this ecosystem, with Google Trends emerging as a powerful, public-facing tool for gaining early insights into evolving patterns. Over the past two decades, Google Trends has been successfully applied in various fields, including epidemiology to predict infectious diseases and outbreaks \citep{yang2015accurate}, macroeconomics to forecast unemployment rates \citep{d2017predictive}, finance to analyze stock market trends \citep{preis2013quantifying}, and marketing to study consumer behavior in fashion \citep{silva2019googling}. One of the earliest and pioneering works in this area, \citep{choi2012predicting}, showed that Google Trends data can enhance real-time estimation of economic indicators, with their 2009 technical report \citep{choi2009predicting} illustrating how search behavior can help ``predict the present''. Compared to traditional data sources such as surveys, these aggregated search volumes are low-cost, continuously updated, and can provide a more dynamic and accurate representation of how behaviors, opinions, and sentiments evolve over time \citep{stephens2017everybody}.

Google Trends provides aggregated search volumes for specific keywords across various geographic scales and time periods. Despite its extensive adoption, the quality of the data presents significant challenges that can affect the reliability of models built on it. Missing values, sampling variability, noise, and unexpected long-term trends arise from the data generation process. For example, privacy-preserving thresholds enforce zero values when searches are too sparse in a region to maintain user anonymity. Daily variations due to Google Trends’ sampling methods introduce noticeable discrepancies across data downloads, which may appear as noise in the reported search volumes. Algorithm updates can affect data consistency over time, resulting in search volumes with different prevalence of zeros, noise levels, and magnitudes that may not reflect meaningful shifts in search patterns. For current and future users of Google trends data, understanding and addressing these issues is critical for unlocking its full potential and enhancing the reliability of models leveraging search data.

\subsection{Outline of the paper}

\begin{figure}[H]
    \centering
    \includegraphics[width=0.4\linewidth]{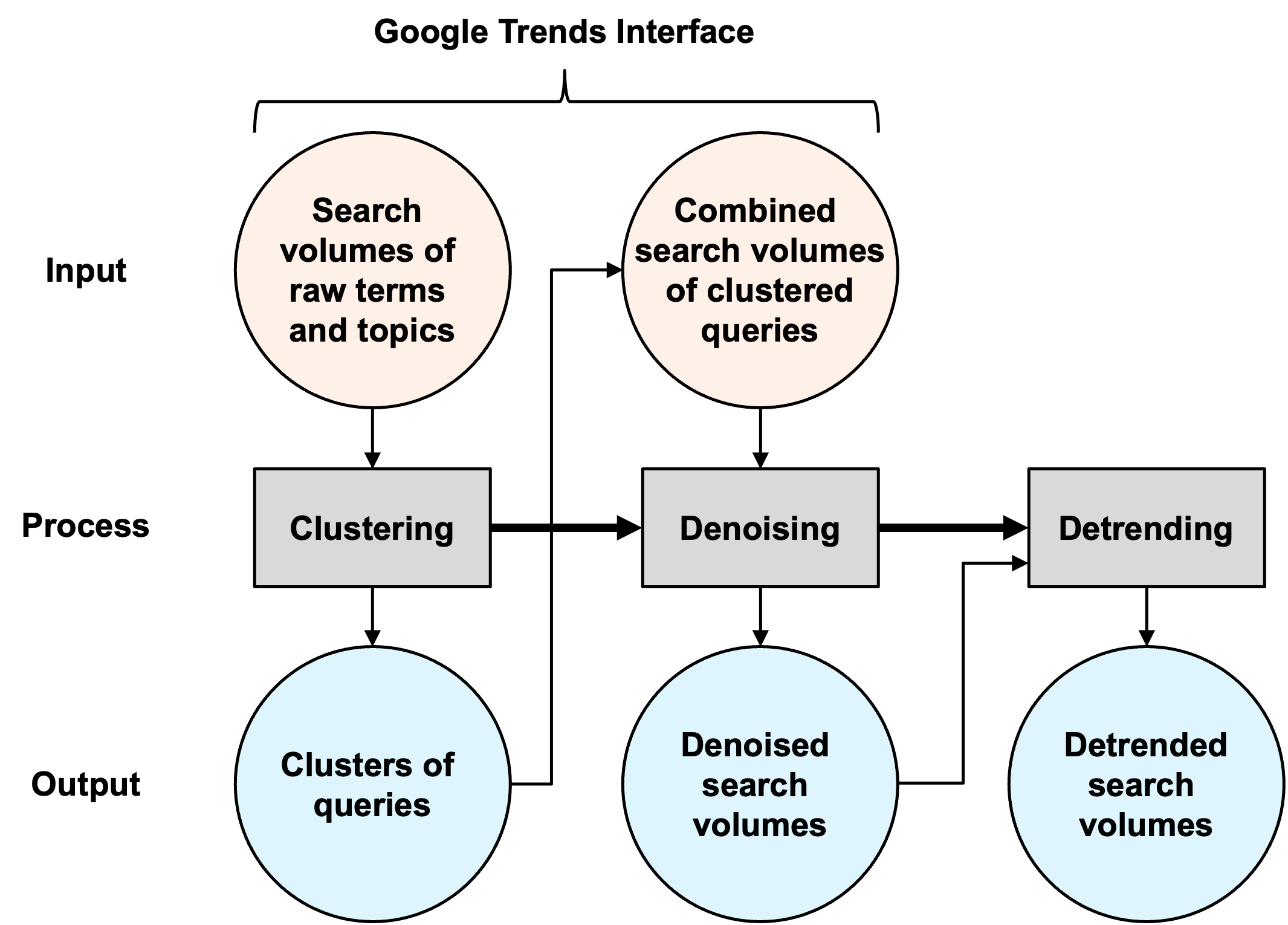}
    \caption{Flowchart of the three-step preprocessing methodology for Google Trends data. Orange circles represent the search volumes retrieved from the Google Trends interface, blue circles represent the outputs produced by each preprocessing step.}
    \label{fig:flowchart}
\end{figure}

In this paper, we present a preprocessing methodology designed to enhance the quality of Google Trends data for real-time forecasting applications. Our approach addresses key challenges inherent in search data, such as missing values, sampling variability, noise, and trends, thereby improving its consistency, stability, and predictive utility.  We begin with a literature review and a dedicated data section providing a practical overview of Google Trends, highlighting its functionalities, challenges, and strategies for selecting relevant search queries. Our statistical methodology is then presented in three stages as illustrated in Figure \ref{fig:flowchart}. (1) First, we employ hierarchical clustering to group similar keywords based on correlations between their time series. Combined search volumes of related terms overcome Google’s privacy thresholds, yielding non-zero values for initially sparse data, thus reducing missing values and stabilizing sampling variability. (2) Next, we apply smoothing splines iteratively to denoise the time series without introducing look-ahead bias, preserving the integrity necessary for real-time forecasting models. This step effectively reduces noise caused by daily sampling variations and short-term fluctuations in searches. (3) To further improve data quality, we use linear detrending, quadratic detrending, and differencing to eliminate long-term deterministic and stochastic trends, ensuring stationarity in the time series while preserving short-term dynamics. We then use correlations between the resulting predictors and the target variables to filter and retain only the most relevant search queries. We validate our preprocessing framework through its application to predicting influenza hospitalizations in the United States using three statistical and two machine learning forecasting models, incorporating Google Trends data as external predictors.

\section{Related Work}
\label{sec:related_work}

Although there is extensive research on the use of Google Trends data for multiple applications, only a few recent analyses have focused on quantifying and addressing data quality issues. Some studies suggest removing all queries whose time series have missing values from the analysis \citep{cebrian2023google, liu2015composite}, resulting in a potential loss of valuable information. To mitigate this, \citet{neumann2023harnessing} recommend combining search phrases in the Google Trends tool using the Boolean operator OR (``+''), which allows the retrieval of aggregated search volumes. However, their approach does not describe how to effectively construct these combinations, leaving a gap in practical implementation.

Sampling variability is another area of concern in Google Trends data. In \citet{cebrian2023google}, the authors highlight that the lack of accuracy in Google Trends data originates from the sampling process, but do not propose specific methods to improve data quality. In contrast, \citet{eichenauer2022obtaining} and \citet{neumann2023harnessing} explain the Google Trends sampling mechanism, which introduces noise and fluctuations in the returned search volumes, and suggest averaging results from multiple downloads to reduce these variations. Expanding on this, \citet{cebrian2024addressing} model and simulate the data generation process to better understand sampling error, proposing a measure to determine the number of samples required for consistent search volumes. However, since Google servers cache query results and return identical values for repeated requests within a 24-hour period, averaging multiple samples may not be practical for real-time analysis.

Several studies have applied temporal smoothing techniques to preprocess Google Trends data before integrating it into forecasting models. For instance, moving averages are employed to smooth time series of search volumes \citep{liu2015composite, rabiolo2021forecasting, wang2022covid, lampos2021tracking}. However, this method can introduce phase shifts, particularly for weekly or monthly data where the peak is delayed after smoothing. Alternatively, \citet{silva2019googling} utilize singular spectrum analysis to denoise search data, demonstrating its effectiveness in enhancing the performance of a neural network autoregression forecasting model. Similarly, \citet{fenga2020filtering} combines a Seasonal AutoRegressive Integrated Moving Average (SARIMA) model with a wavelet denoising filter, showing that the addition of the filter improves forecasting accuracy compared to using SARIMA alone. While these studies effectively address noisy signals and underscore the benefits of denoising Google Trends data to enhance forecasts, their methodologies rely on the entire time series for smoothing, including future time periods. This makes them unsuitable for real-time forecasting, as future data is not available at the time of prediction. Retrospective application of these techniques could introduce future-looking information, potentially leading to unrealistic accuracy improvements not achievable in real-time settings.

Various online blog case studies also explore methods for analyzing and potentially enhancing Google Trends data. For instance, \citet{matsa2017flint} suggest retrieving search volumes of terms grouped into broad categories using domain knowledge, imputing the remaining zeros with regression, and smoothing fluctuations with a generalized additive model. Their methodology, while comprehensive, lacks reproducibility due to the manual selection and combination of keywords and the need for multiple Google accounts to circumvent data download limits.

Clustering semantically related terms as a preprocessing step has also been explored. \cite{ning2024incorporating} group semantically related search queries using hierarchical clustering based on correlation distance. A group-wise L2 penalty is applied to these clusters, allowing the model to retain or eliminate entire sets of terms. However, while clustering queries that share meaning is beneficial, the group-wise penalty may inadvertently discard individual terms within clusters that have strong predictive power, potentially overlooking valuable information. Similarly, \citet{lampos2015advances} use k-means clustering with cosine similarity to group queries that share semantic and temporal patterns.  However, the number of clusters for k-means needs to be specified in advance, which can be challenging without first analyzing the cluster structure.

One of the earliest and most significant applications of Google Trends has been infectious disease forecasting in the United States (US). Influenza prediction is a well-studied application of Google Trends data and therefore serves as the case study to demonstrate the effectiveness of our preprocessing methodology. The first documented use of Google search data was presented by \citet{ginsberg2009detecting}, who leveraged search volumes to forecast influenza outbreaks, leading to the development of Google Flu Trends (GFT) tool. Following the failure of GFT to provide reliable forecasts \citep{santillana2014can}, substantial research has been conducted on refining Google Trends-based predictions by integrating data from multiple sources and calibrating machine learning models \citep{yang2017using, kandula2019improved, ning2019accurate, yang2021use}. In particular, \citet{yang2015accurate} improved the GFT framework with ARGO (AutoRegression with GOogle search data), an autoregressive model that selects the most relevant exogenous predictors to accurately estimate influenza-like illness (ILI) rates based on historical flu activity and Google search data. Building on this, \citet{ning2019accurate} developed ARGO2 (2-step Augmented Regression with GOogle data), which refines ARGO by combining national and regional predictions, incorporating cross-regional correlations, and computing the final ILI estimates using the best linear predictor. The performance of these models shows the benefits of integrating Google Trends data in predictive models applied to influenza forecasting. 

\subsection{Our Contributions}
\label{sec:contributions}

Our contributions are threefold. (1) We present a general preprocessing methodology that enhances the quality of Google Trends data to improve real-time forecasting performance. (2) Through the proposed approach, we address multiple challenges inherent in search data, including missing values, sampling variability, noise, and long-term trends. The framework is fully reproducible and designed for practical implementation. It consists of a three-step procedure that transforms raw search volumes into reliable and informative features while preserving the integrity required for real-time settings. (3) We apply our methods to influenza hospitalization forecasting across diverse statistical and machine learning models. Our results show that using preprocessed search data significantly increases accuracy compared to omitting external variables, while using raw data degrades performance in statistical models. Although our case study focuses on infectious disease, the proposed approach is broadly applicable to other domains where Google Trends data can be used as an exogenous indicator.

\section{Data}
\label{sec:data}

Google Trends is the primary data source for this study, providing search volumes for flu-related terms. Available publicly at \url{https://trends.google.com/trends/}, it offers valuable insights into search behaviors but presents challenges due to its data generation mechanism. 

\subsection{Description of Google Trends}
\label{sec:description}

Google Trends allows users to analyze search behavior for inputted keywords by location, time range, category and search type. The search volume is available for two types of keywords: ``search terms'' and ``topics''. A search term can be a single word or a phrase, with returned volumes encompassing searches that include the words in any order, as well as with additional words before or after. Boolean logic can be used to retrieve specific phrases or exclude some terms \citep{trends-tips}. Topics are broader than search terms and typically yield higher search volumes. They are ``generally considered to be more reliable for Google Trends data'' \citep{trends-basics}, as they account for all languages and spellings of a word. Moreover, results for topics are more refined and less noisy, as these are associated with a category (such as ``Health'' or ``Finance''), which returns results in the desired field when a term has multiple meanings \citep{trends-categories}. 

In fact, the tool allows to refine the search to only include results for a particular area of interest or ``category'', rather than including all searches. The category then works as a filter for the data, to exclude irrelevant searches. As another filter, there are different ``search types'', e.g., web search or Youtube search. In this paper, we analyze web searches, which is the case of most studies. Results can be obtained for any location, ranging from worldwide to a particular country, state, metro, or city. Data is available for any time range from 2004 to the present. 

\begin{figure}[h]
    \centering
    \includegraphics[width=0.6\linewidth]{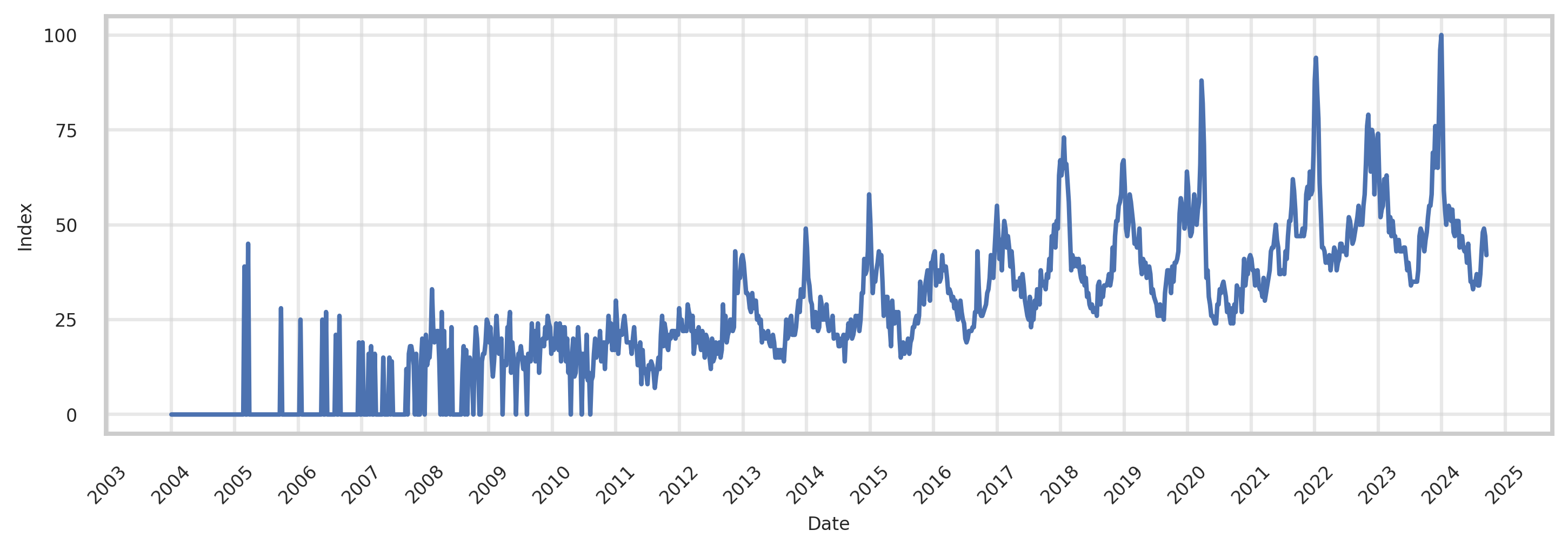}
    \caption{Example of search volumes over time for ``Cough'' in Alabama}
    \label{fig:Example_volumes}
\end{figure}

For a specific search, there are four types of data representations: ``interest over time'', ``interest by subregion'',``related queries'' and ``related topics''. The ``interest over time'' feature returns a time series of search volumes that reflects the relative popularity of a keyword over time, scaled from 0 to 100, where 100 corresponds to the peak volume \citep{trends-faq}. This time series reveals trends, seasonal patterns, or spikes in interest, and can be obtained at various temporal resolutions, e.g., hourly, daily, weekly or monthly. Figure \ref{fig:Example_volumes} displays a representative example of search volumes for the keyword ``Cough''. ``Interest by subregion'' provides a geographical breakdown  of search data, highlighting the areas where a subject is most popular. By explicitly selecting a location in the input box, the interest over time also provides the interest in that particular region. 

To further explore relevant search activity, the ``related queries'' and ``related topics'' functionalities help us identify a broader set of terms and topics connected to the initial search keyword. They can be selected as ``rising'', that is, emerging terms that had a significant increase in search volume, or ``top'', that is, most popular search terms in the period, location and category chosen \citep{trends-related}. Hence, choosing different time periods, regions or categories will give different related searches. Section \ref{sec:flu_terms} provides a more detailed approach to obtaining relevant keywords.

Google Trends provides real-time data that offers up-to-date information on search trends. This includes partial data retrieval during ongoing periods, such as the middle of the day, week or month, which is valuable for real-time forecasting. For example, the start of the week for weekly search volumes is Sunday, which allows to obtain partial volumes on Wednesday for the first four days of the week. 

On the software development side, there are different ways to retrieve Google Trends search volumes. A csv file can be downloaded directly from the website. There exist \texttt{R} and \texttt{Python} packages (\texttt{gtrendsR} and \texttt{pytrends}) for convenient download. Google Trends contains an Application Programming Interface, which gives identical results to the Google Trends publicly available site but automates the process of downloading the data. For this paper, all our methods are applicable to direct downloads of search volumes from \url{https://trends.google.com/trends/}.

\subsection{Challenges of Google Trends Data}
\label{sec:challenges}

\begin{figure}[h]
    \centering
    \begin{subfigure}[b]{0.3\linewidth}
        \centering
        \includegraphics[width=\linewidth]{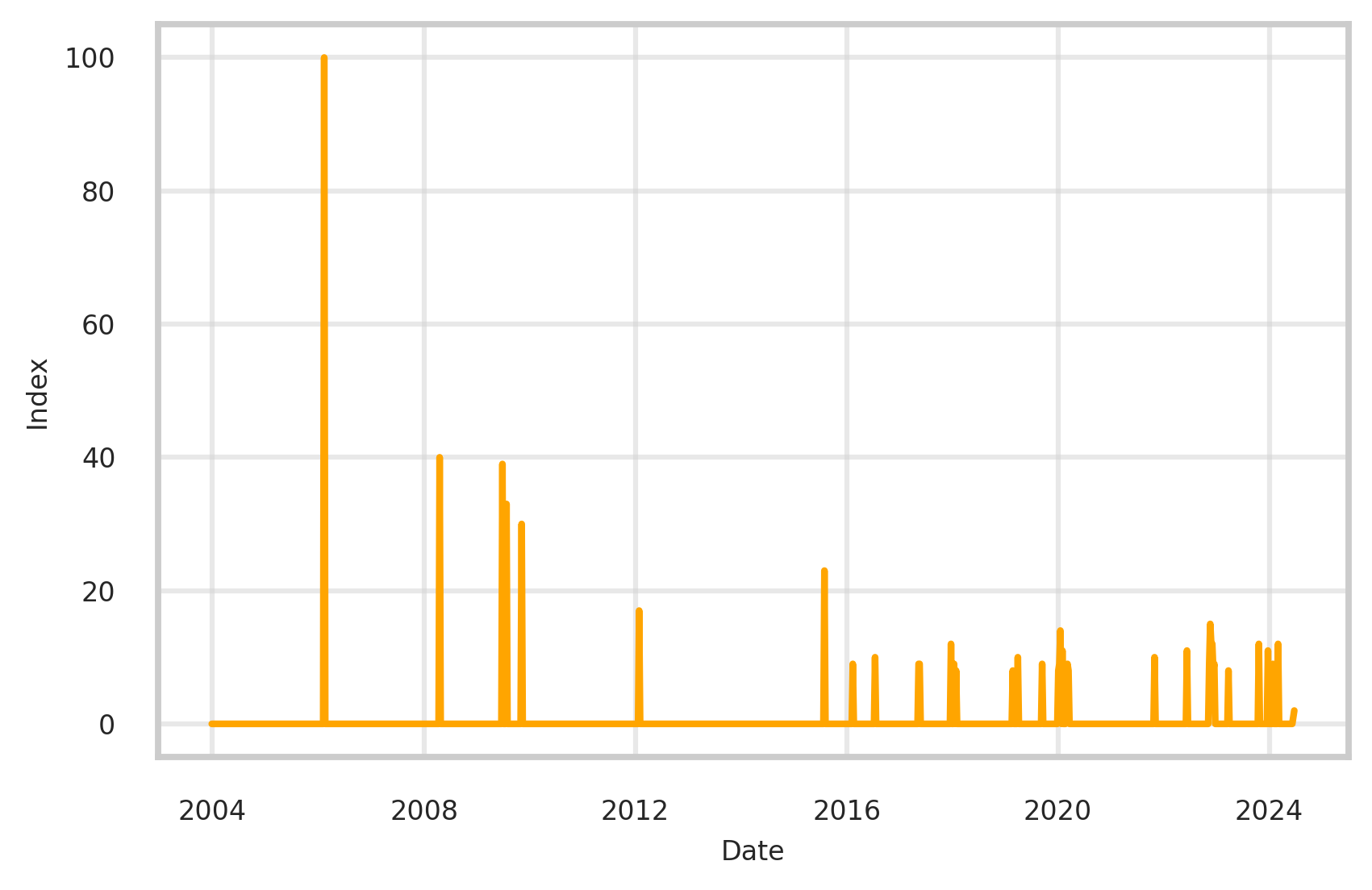}
        \caption{Missing Values in ``Influenza Treatment''}
        \label{fig:missing_values}
    \end{subfigure}
    \hfill
    \begin{subfigure}[b]{0.3\linewidth}
        \centering
        \includegraphics[width=\linewidth]{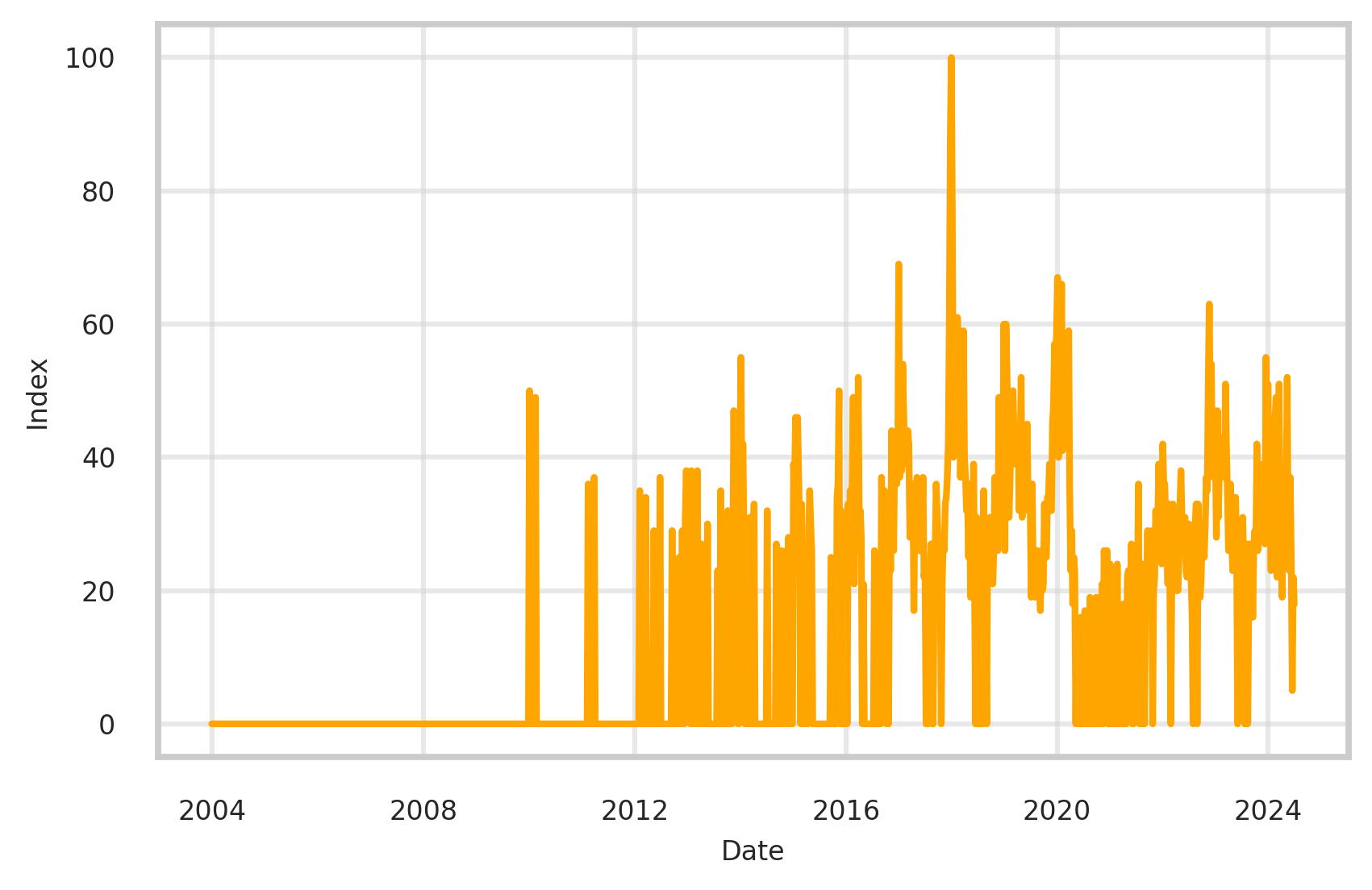}
        \caption{Noise in ``Acute Bronchitis'' \vspace{4mm}}
        \label{fig:noise}
    \end{subfigure}
    \hfill
    \begin{subfigure}[b]{0.3\linewidth}
        \centering
        \includegraphics[width=\linewidth]{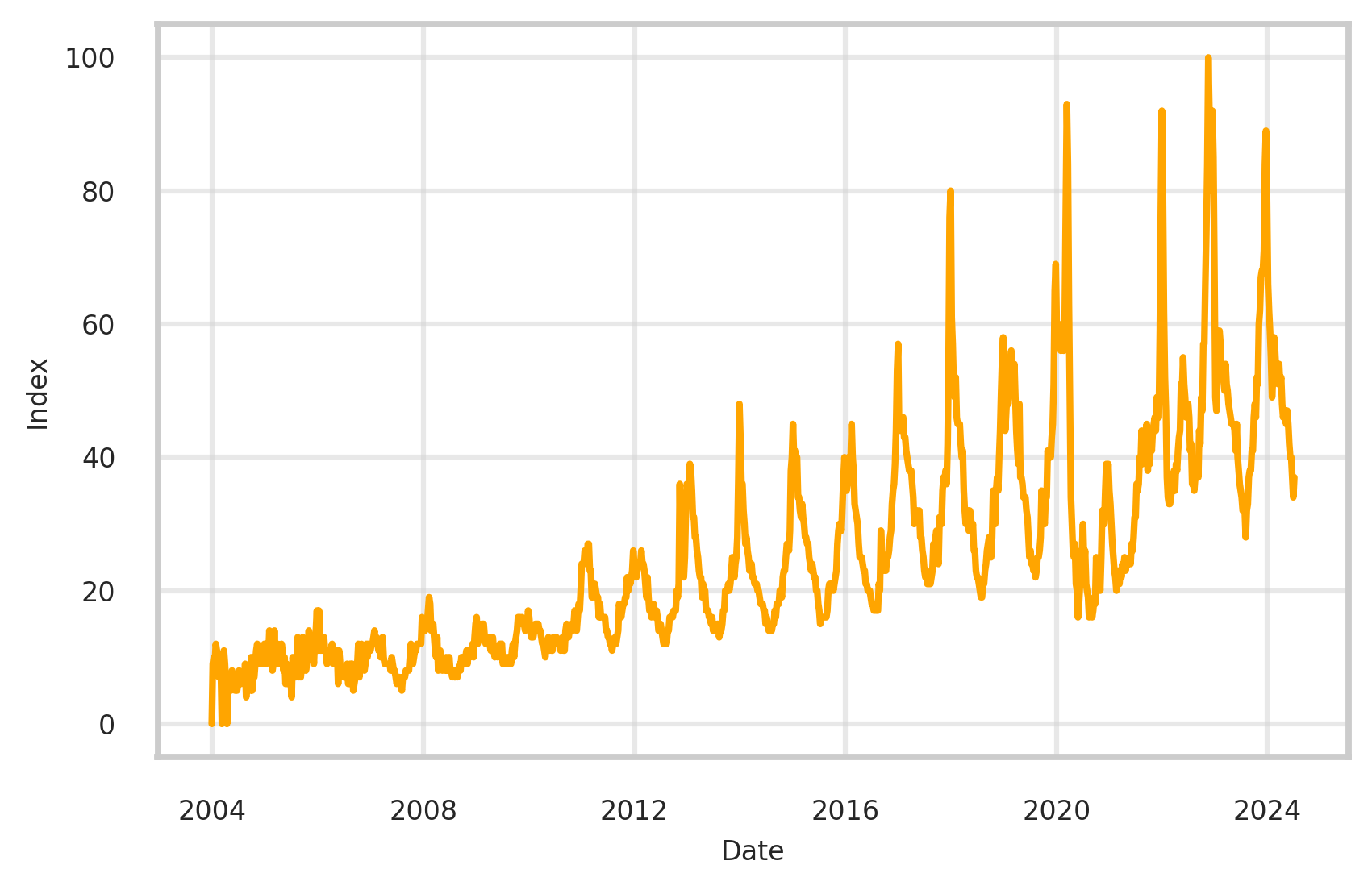}
        \caption{Trend in ``Cough'' \vspace{4mm}}
        \label{fig:trend}
    \end{subfigure}
    \begin{subfigure}[b]{\linewidth}
        \centering
        \includegraphics[width=0.5\linewidth]{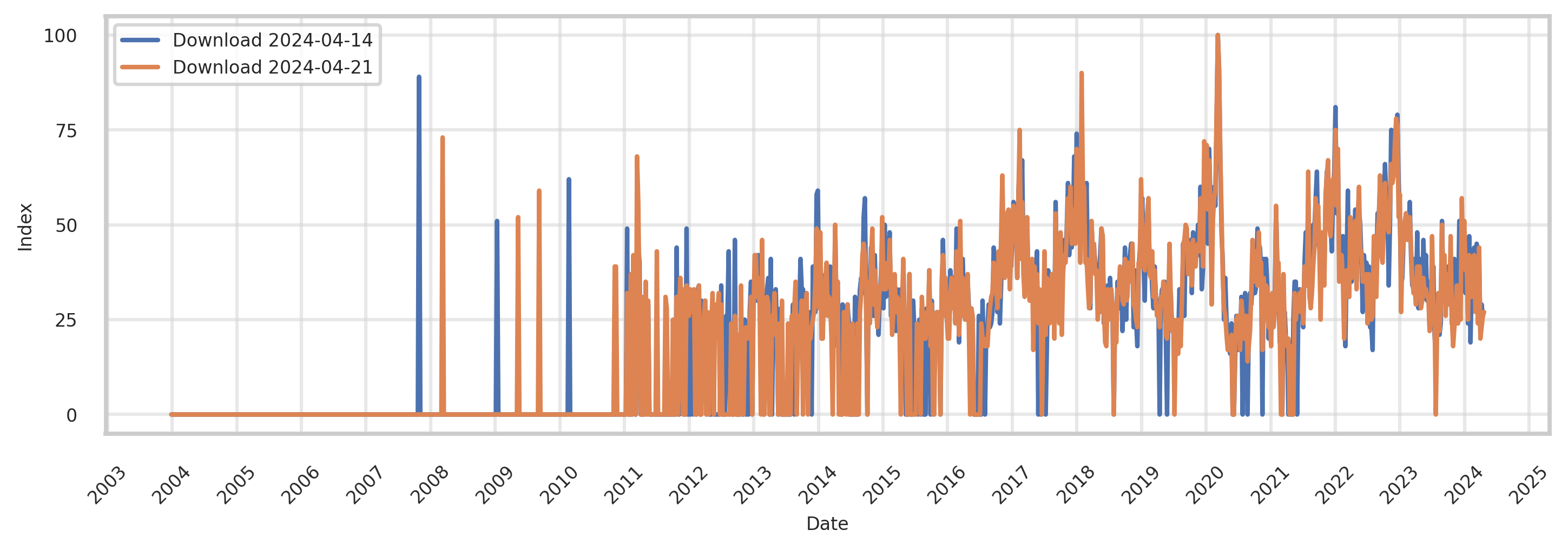}
        \caption{Sampling Variability across different downloads for ``Common Cold''}
        \label{fig:variability}
    \end{subfigure}
    \caption{Examples of issues in Google Trends search volumes for California and Alaska}
    \label{fig:search_issues}
\end{figure}

The data generation mechanism in Google Trends  introduces issues such as missing values, sampling variability, noise, and trends, that can affect forecasting quality. Figure \ref{fig:search_issues} illustrates these challenges using search volumes for four different terms. 

Google Trends assigns a value of ``0'' when the number of searches for a specific keyword falls below undisclosed privacy thresholds defined by Google, as illustrated in Figure \ref{fig:missing_values}. It indicates low search activity rather than a complete absence of searches. This can result in sparse data, particularly at the state or city level and for daily or weekly frequencies \citep{stephens2014hands}. These zeros represent missing data that need to be addressed, as they mask low but potentially meaningful search activity.

Noise, shown in Figure \ref{fig:noise}, is also inherent in the data generation process of Google Trends.  Given the billions of Google searches conducted daily, Google Trends provides only a representative sample to ensure quick response times \citep{trends-faq}. This sample is a random approximation of the entire population of searches, and thus contains sampling noise. To optimize processing, Google servers cache data for 24 hours, meaning that repeated input searches within the same day yield identical results. However, at midnight UTC, the cache resets, and a new sample is drawn, leading to day-to-day variability across different downloads of search volumes for the same search \citep{choi2012predicting, neumann2023harnessing}, as illustrated in Figure \ref{fig:variability}. This variability arises from the randomness of the sampling-based data generation process, which follows a hypergeometric distribution \citep{bleher2022knitting}. Let $K$ and $k$ be the search volumes for a keyword in a population of size $N$ and a sample of size $n$, respectively. Since $K$ and $N$ are unobserved, the reported proportion $\frac{k}{n}$ is a noisy approximation of $\frac{K}{N}$. Google Trends scales search volumes from 0 to 100, effectively returning $100 \times \frac{k}{n}/\frac{K}{N}$. As $K$ decreases, the variance of $\frac{k}{n}/\frac{K}{N}$, which reflects sampling variability, increases. This leads to noisier estimates, especially for small regions and infrequent searches.

As terms become more or less popular with rising or declining public interest, search volumes may exhibit a trend, as displayed in Figure \ref{fig:trend}. These trends represent gradual changes over time that can mask underlying patterns. In the context of Google Trends, this phenomenon can have multiple causes, including changes in vocabulary with new or obsolete terms, shifts in general public interest in a subject, or the platform's dynamic normalization. Specifically, Google Trends scales its ``interest over time'' index from 0 to 100 within the selected time period and region. When the temporal or spatial window changes, the entire series is rescaled so that the new maximum equals 100, and all other values are adjusted proportionally. This can introduce artificial level shifts, low-frequency drifts, and apparent structural breaks that resemble long-term trends but do not reflect actual changes in search behavior. Such spurious trends can bias correlations and obscure long-term relationships with the target variable.

Additionally, data quality could change as Google Trends' algorithm undergoes updates. \citet{cebrian2024addressing} mention that the platform has undergone notable algorithmic updates over the years. According to the Google Trends website, improvements to geographical assignment and data collection systems were implemented in 2011, 2016, and 2022, as shown in SI Figure \ref{fig:gt_notes}. These updates can lead to inconsistencies. For example, \citet{myburgh2022infodemiologists} reports that changes to Google's sampling strategy in January 2022 resulted in higher search volumes compared to earlier data. Consequently, data after January 2022 is not directly comparable to prior periods. Therefore, long-term trends may also reflect algorithmic changes rather than genuine shifts in search behavior, posing challenges for accurate forecasting. 

\begin{figure}[H]
    \centering
    \includegraphics[width=0.7\linewidth]{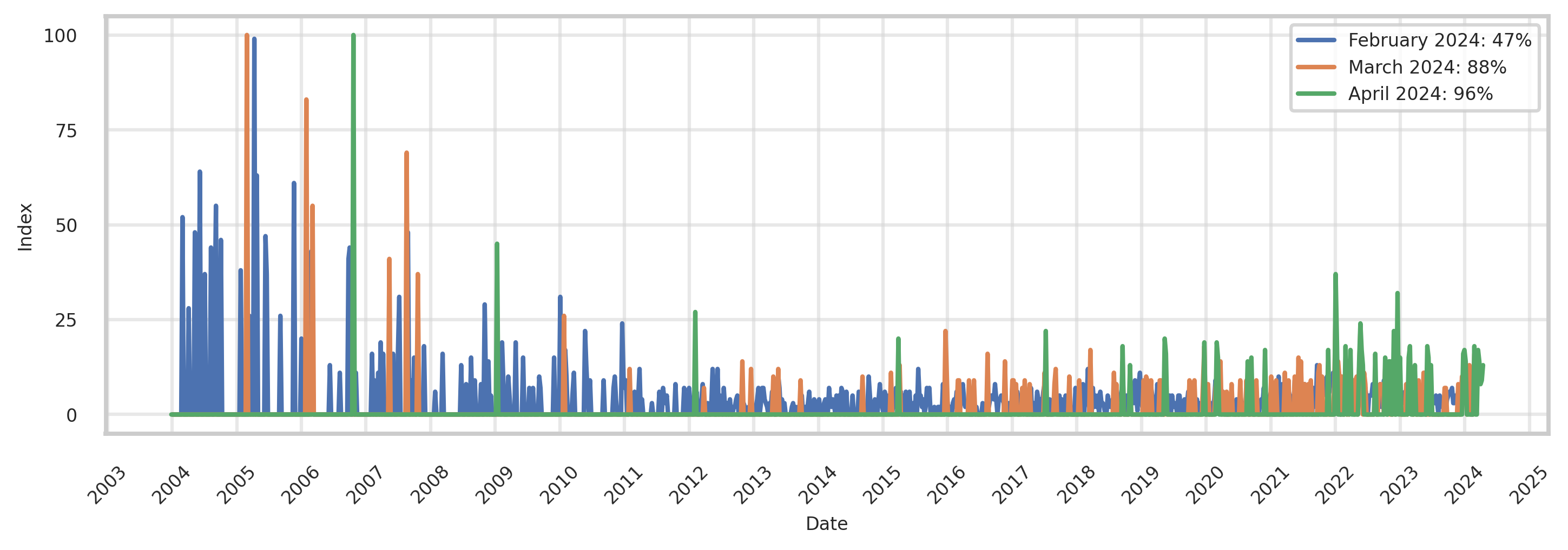}
    \caption{Example of changes in the percentage of zeros within different downloads of ``Nasal Congestion'' in Alaska from February to April 2024}
    \label{fig:updates}
\end{figure}

We have also observed changes in Google Trends data within our dataset, where the total number of zeros increased by 40\% at the start of 2024 (See SI Table \ref{tab:zeros_updates}). Figure \ref{fig:updates} shows a 104\% increase in the proportion of zeros for keyword ``Nasal Congestion'' between February and April 2024. This increase may be linked to another algorithm update, as the elevated number of zeros has been constant since. Therefore, it is critical to construct a preprocessing strategy that can handle updates from the data collection system.

\section{Methodology}
\label{sec:methods}

In this section, we present our methodology to handle missing values, sampling variability, noise, and trends in Google Trends' search volumes using hierarchical clustering, smoothing splines, and detrending techniques. 

\subsection{Missing Values and Sampling Variability}
\label{sec:clustering}

Zeros in Google Trends data, due to low search volumes, represent missing values. Addressing this issue is essential, as even minimal search activity can contain valuable insights. A common approach is to discard affected terms, but it risks eliminating meaningful information. An alternative is to impute zeros, which can introduce errors, especially in time series with high sparsity, as there may not be enough information to produce accurate estimates. To overcome this issue, we implement a hybrid approach, by discarding terms with insufficient data (over 99\% of zeros), using keywords with fewer than 30\% of missing values as individual predictors, and combining the remaining queries whose aggregated search volumes exceed privacy thresholds. Moreover, since phrases can be searched in any word order, we remove duplicated terms (e.g. ``cold and flu'' and ``flu and cold'') by computing pairwise correlations between time series, and discarding one of the two variables when their correlations are higher than 0.99.

Google Trends enables the use of Boolean logic, where the ``+'' operator is interpreted as ``or'', to obtain the union of search activity for multiple terms. This results in higher overall volumes with fewer zeros. As categorizing terms using expert knowledge is impractical and lacks reproducibility, we use a data-driven approach to ensure a systematic grouping. We associate terms using hierarchical clustering based on correlations as a similarity metric, following \citet{ning2024incorporating}'s recommendation.

Hierarchical clustering \citep{ward1963hierarchical} is a data analysis technique that creates a sequence of nested clusters based on a hierarchical structure \citep{maharaj2019time}. We use Ward's method to determine how clusters are merged, as it minimizes the total within-cluster variance at each step. This approach produces spherical clusters of relatively similar sizes \citep{maharaj2019time, everitt2011cluster}, ensures clear separation, and is regarded as one of the most robust and effective methods for noisy data \citep{balcan2014robust}. Clusters with high similarity are merged together in successive steps. In hierarchical clustering, smaller, closely related clusters can be combined into broader groups, while larger ones can be subdivided to enhance relevance and specificity. 

The similarity metric used in our hierarchical clustering is correlation distance, which is defined as $(1-\text{correlation matrix})$. Since different phrasings of a query often produce similar search results, their search volumes tend to align as well. Correlation matches time series that share similar shapes and have synchronized peaks and troughs. It also enables sparse signals with relevant spikes to be grouped with more complete ones, provided their key patterns align. Clustering keywords based on correlations thus forms groups that not only share semantic meanings but also have combined search volumes that surpass Google's privacy thresholds. This approach effectively addresses the challenge of zeros while preserving meaningful search activity.

To determine the optimal number of clusters for hierarchical clustering, we use the elbow method, which plots the within-cluster sum of squares (WCSS) against the number of clusters. WCSS, defined as $\sum_{i=1}^{k} \sum_{x \in C_i} |x - \mu_i|^2$, where $\mu_i$ is the average time series of cluster $C_i$, measures the sum of squared distances between each time series and its cluster center \citep{maharaj2019time, hastie2017elements}. As the number of clusters increases, WCSS decreases, with the optimal point occurring at the ``elbow,'' where adding more clusters yields diminishing returns. At the elbow, clusters strike a balance between capturing meaningful patterns and maintaining a sufficient size to surpass Google’s privacy thresholds. If a single cluster remains disproportionately large compared to the others, we apply a second round of hierarchical clustering to further subdivide it. The categorized queries are then grouped using the Boolean ``+'' operator, and their combined search volumes are retrieved from Google Trends. We finalize the clusters after confirming they contain low percentages of zeros. 

Clustering not only addresses missing values, but also sampling variability. Popular keywords with high search volumes exhibit consistent behavior across different samples, while less popular terms with sparse data are subject to higher sampling variability \citep{medeiros2021proper}. Grouping such queries into clusters increases search volumes, effectively reducing hypergeometric sampling randomness and improving data reliability.

\subsection{Noise}
\label{sec:noise}

Noise in Google Trends data arises from random sampling variations, which can obscure the underlying signal and lead to overfitting. While various statistical techniques can estimate the true signal, real-time forecasting requires smoothing time series using only past data to prevent information leakage, where data not available at the time of estimation is inadvertently included in the model \citep{joseph2022modern, borup2022search}. Exponential smoothing and moving averages rely on past values but may introduce phase shifting, while K-nearest neighbors, kernel smoothing, and LOESS require both past and future points. Polynomial regression risks overfitting or underfitting, and Fourier transforms and wavelets need the entire time series for frequency and time-frequency analysis \citep{shumway2006time, brockwell2016introduction}. Such constraints make these methods unsuitable for real-time applications. To overcome these challenges, we apply smoothing splines iteratively within a rolling window of past values. This approach offers localized data fitting, relies exclusively on past observations, provides precise control over smoothing, and has been shown to effectively remove noise \citep{musial2011comparing, navarrete2018prediction}. 

Smoothing splines \citep{reinsch1967smoothing} aim to minimize the function 
\begin{equation}
    \sum_{t=1}^{n} (x_t - f_t)^2 + \lambda \int (f''_t)^2 dt,
\end{equation}
where $x_t$ are the observations, $f_t$ is the cubic spline, and $\lambda$ is the smoothing parameter controlling the degree of smoothness \citep{hastie2017elements, shumway2006time}. Larger $\lambda$ values produce smoother time series but risk shrinking peaks and flattening the signal, while smaller values may retain excessive noise. To balance these effects and avoid over-smoothing, we restrict $\lambda$ to the range [0.1, 2]. 

We implement an iterative denoising process tailored for real-time settings, performing one-step-ahead predictions of Google Trends search volumes. The data is split into training and testing sets, with optimal $\lambda$ values for each keyword identified through a grid search on the training set, and the test set used for out-of-sample denoising. During training, we fit smoothing splines on a rolling window of recent observations, and the next data point outside the window is predicted. The window then advances by one observation, and the process is repeated across the training set. During testing, smoothing splines are applied similarly, but the last fitted observation within the rolling window, not the predicted one, is used as the denoised value. This approach prevents overfitting to recent trends and ensures smoother transitions.

Optimal $\lambda$ parameters are selected by minimizing the Root Mean Squared Error (RMSE) between the one-step-ahead predictions and the raw values. RMSE captures peaks and large variations common in Google Trends data, which often exhibits sudden spikes in search volumes. By penalizing large differences, this metric ensures that smoothing preserves peak magnitudes without excessive flattening. 

The rolling window length is chosen based on the characteristics of the data and domain knowledge. According to the CDC, an influenza season typically spans from October to May, or roughly 30 weeks \citep{cdchosp, cdcfluseason}. To avoid smoothing across multiple seasons and diluting within-season dynamics, the window should not exceed this activity period. We use a 20-week window, which represents about 70\% of the season length, to ensure that both seasonal patterns and sudden spikes are retained. If the window is too short, denoising may not effectively reduce noise, while if it is too long, denoising may attenuate meaningful short-term variations.

Denoising is applied only to time series with high noise levels, identified by evaluating the training RMSE associated with the optimal $\lambda$. Higher RMSE values indicate that noise dominates the series, making it difficult for the model to produce accurate predictions. To determine which series to denoise, we set a threshold based on the empirical distribution of RMSEs across all time series. Specifically, series with RMSE greater than the median RMSE are considered noisy and are smoothed. This strategy allows us to selectively improve noisy series while avoiding unnecessary adjustments for cleaner ones.

\subsection{Trend}
\label{sec:trend}

Trends in Google Trends data arise from long-term changes in search volumes, often driven by shifts in public interest, dynamic normalization, or algorithm updates. While most keywords have search volumes with a relatively stable mean over time, some exhibit persistent trends. However, for applications such as predicting infectious disease surges, unemployment rates, or stock prices, the focus is typically on short-term patterns. Long-term trends can obscure these dynamics and introduce noise into forecasting models. Furthermore, a time series with a trend is non-stationary, complicating the accurate estimation of autocorrelations between observations \citep{shumway2006time} and potentially distorting variable relationships. Detrending therefore enhances the interpretability of the data and supports more robust comparisons between search interest and target outcomes. We apply methods such as model fitting or differencing, to remove trends and restore stationarity. Having access to Google Trends data since 2004 provides an advantage, allowing for a clear estimation of the long-term trend. 

Time series trends can be classified as either deterministic or stochastic. A deterministic trend is a fixed and predictable pattern over time, following a linear or quadratic shape, and can be removed through detrending to achieve stationarity. A stochastic trend, on the other hand, arises from the cumulative effect of past values and random shocks, and should be eliminated through differencing. To identify the presence and the type of trend in a time series, we apply the Augmented Dickey-Fuller (ADF) test \citep{dickey1979distribution} to each search volume time series. The null hypothesis ($H_0$) states that the time series has a unit root and is non-stationary, while the alternative hypothesis ($H_1$) is that the time series is stationary, potentially around a deterministic trend \citep{joseph2022modern}. Since search volumes are strictly non-negative, their time series have a non-zero mean. The ADF test fits a regression model to the time series data, as follows \citep{valenzuela2019theory}:
\begin{equation}
    \label{adf_eq}
    \Delta x_t = \mu + \alpha t + \beta t^2 + \gamma x_{t-1} + \sum_{i=1}^{p-1} \phi_i \Delta x_{t-i} + \epsilon_t,
\end{equation}
where $x_t$ is a time series, $\Delta x_t = x_t - x_{t-1}$ is the first difference, $\mu$ is the mean, $\alpha$ and $\beta$ are coefficients for linear and quadratic time trends, $t$ is the time index, $\gamma$ determines stationarity, $p$ is the lag order, and $\epsilon_t$ is the error term.

The ADF test operates by estimating the regression equation \eqref{adf_eq} to evaluate stationarity and guide necessary transformations under different assumptions about the type of trend in the data. We test $H_0 : \gamma = 0$ against $H_1: \gamma < 0$. We first assess whether the series is stationary around a constant, non-zero mean, under the assumption $\mu \neq 0, \alpha = 0, \beta = 0$. If $H_0$ is rejected under this assumption, no further transformation is necessary. If not, the series may contain a trend. Next, we test whether the series is stationary around a linear deterministic trend, under the assumption $\mu \neq 0, \alpha \neq 0, \beta = 0$. If $H_0$ is rejected under this assumption, the series requires linear detrending. Otherwise, it may contain a quadratic or stochastic trend. Finally, we test whether the series is stationary around a quadratic deterministic trend under the assumption $\mu \neq 0, \alpha \neq 0, \beta \neq 0$. Rejection of $H_0$ under this assumption indicates the need for quadratic detrending. If all tests fail to reject $H_0$, the series exhibits a stochastic trend and requires differencing. We use the significance level of $\alpha = 0.05$.

To perform detrending, we estimate deterministic trends using ordinary least squares (OLS) regression and subtract the fitted line from the data. A linear trend is estimated as $\hat{m}_t = \mu + \alpha t$, while a quadratic trend is modeled as $\hat{m}_t = \mu + \alpha t + \beta t^2$. We remove the trend from the data as $x_t - \hat{m}_t$. Differencing, used for stochastic trends, is computed as $\hat{x}_t = x_t - x_{t-1}$. To avoid look-ahead bias, we estimate trend parameters using only the training set and apply transformations to the entire time series. Only the test set is used for model evaluation. 

\section{Results}
\label{sec:results}

This section presents a modularized evaluation of each preprocessing step, including clustering, denoising, and detrending, applied to Google Trends flu-related keywords, and highlights their individual contributions to improving data quality. 

\subsection{Evaluation of Clustering}
\label{sec:clustering_results}

Hierarchical clustering is applied to all terms and topics in the candidate pool of 500 terms for each location to produce clusters of keywords with denser time series. We evaluate the clustering results by visualizing search volumes and analyzing the percentage of zeros.

\begin{figure}[H]
    \centering
    \begin{subfigure}[b]{\linewidth}
        \centering
        \includegraphics[width=0.7\linewidth]{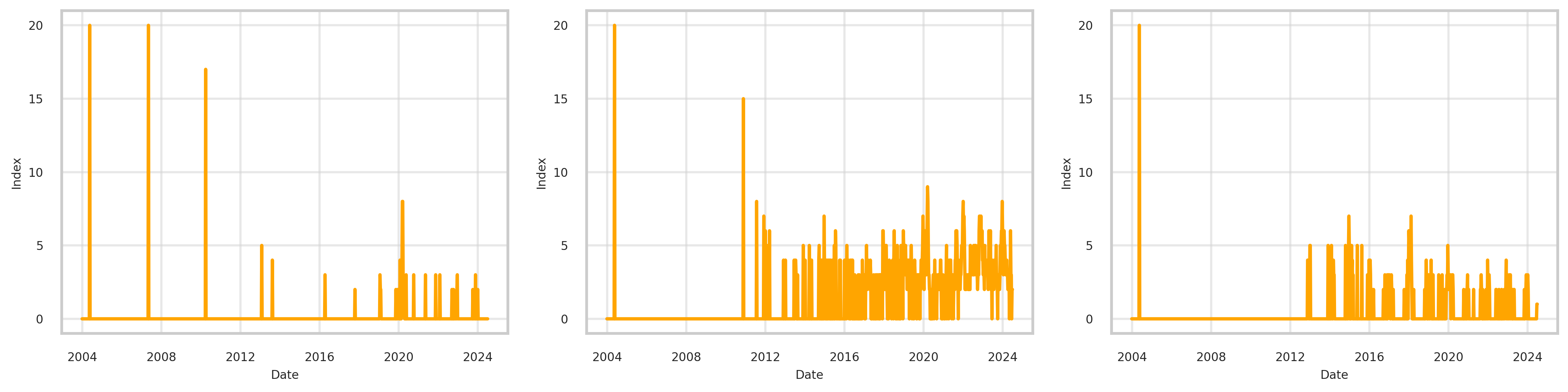}
        \caption{Individual (large values truncated)}
        \label{fig:indiv}
    \end{subfigure}
    \begin{subfigure}[b]{0.4\linewidth}
        \centering
        \includegraphics[width=\linewidth]{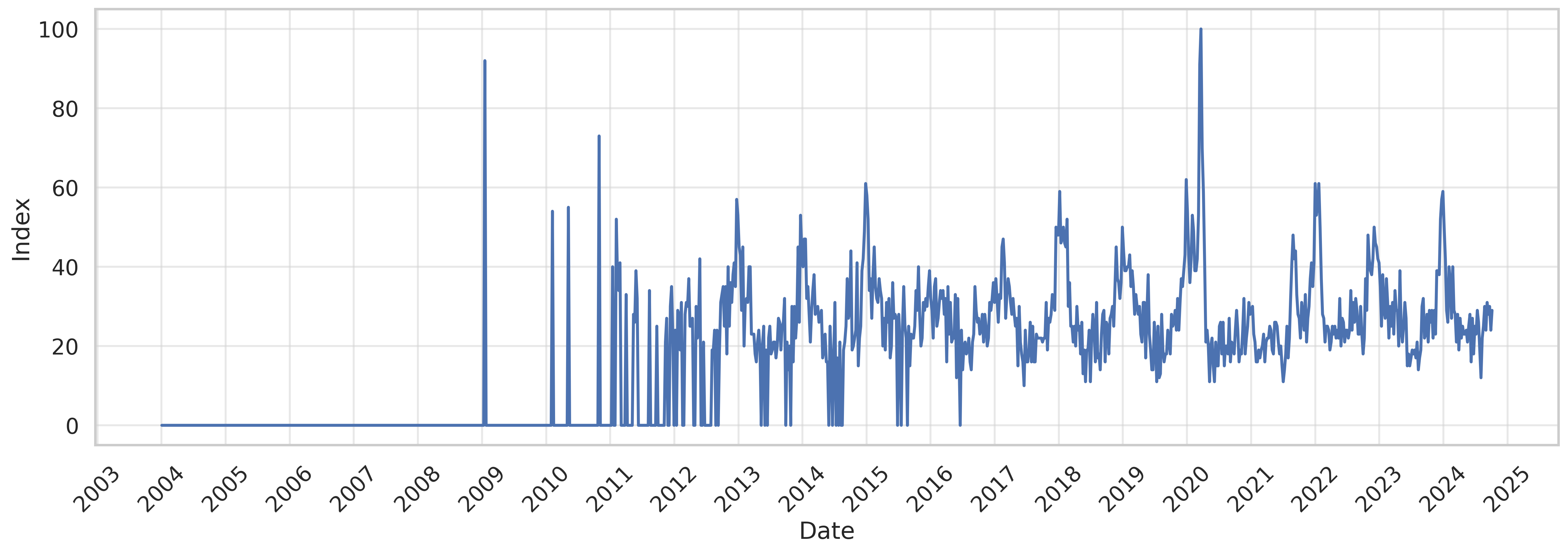}
        \caption{Combined}
        \label{fig:combined}
    \end{subfigure}
    \begin{subfigure}[b]{0.4\linewidth}
        \centering
        \includegraphics[width=\linewidth]{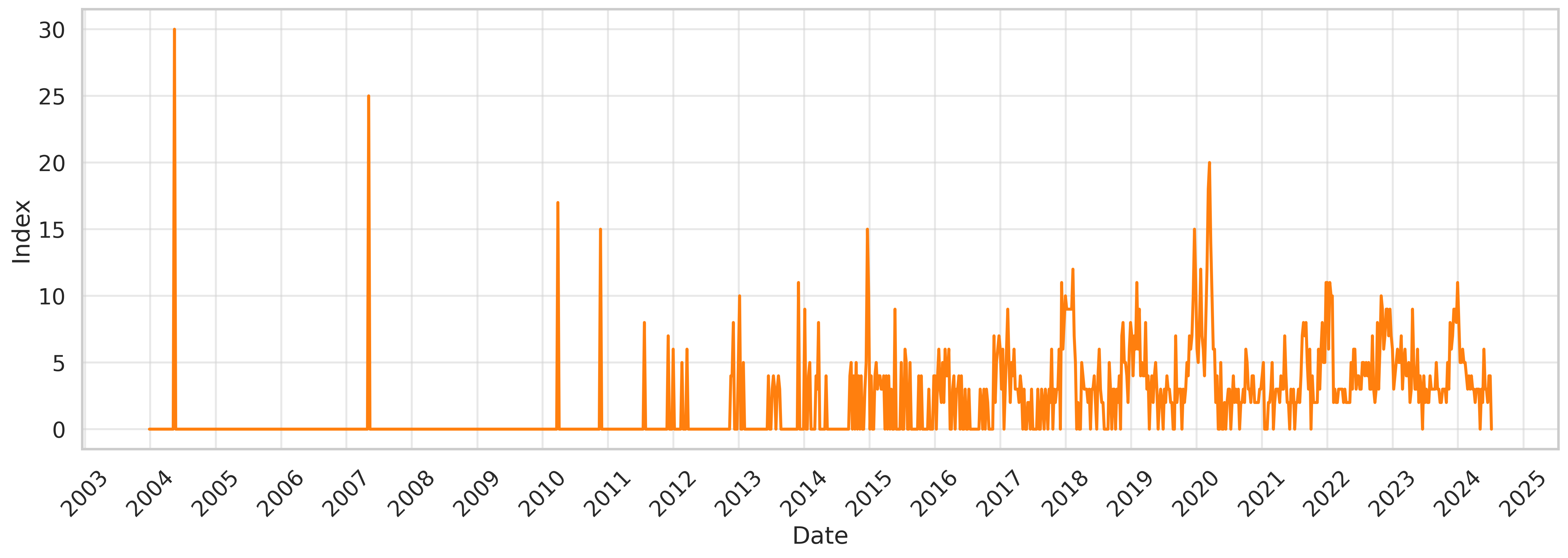}
        \caption{Summation (large values truncated)}
        \label{fig:sum}
    \end{subfigure}
    \caption{Example comparison of search volumes in Tennessee}
    \label{fig:indiv_comb_sum}
\end{figure}

Figure \ref{fig:indiv_comb_sum} compares the raw, combined and summed search volumes for an example cluster for Tennessee, composed of three keywords (``cold virus'', ``high fever'', ``symptoms of pneumonia'') whose individual time series displayed in Figure \ref{fig:indiv} have 96\%, 63\%, and 86\% of zeros. Figure \ref{fig:combined} reveals that their combined search volumes have a reduced percentage of zeros (40\%) with clear and interpretable patterns. In contrast, simply summing the individual time series, as shown in Figure \ref{fig:sum}, results in 60\% of zeros and less discernible patterns. This illustrates that clustering is an effective method for addressing data sparsity, as it generates usable time series by aggregating sparse keywords into meaningful groups. 

\begin{figure}[H]
    \centering
    \begin{subfigure}[b]{0.3\linewidth}
        \centering
        \includegraphics[width=\linewidth]{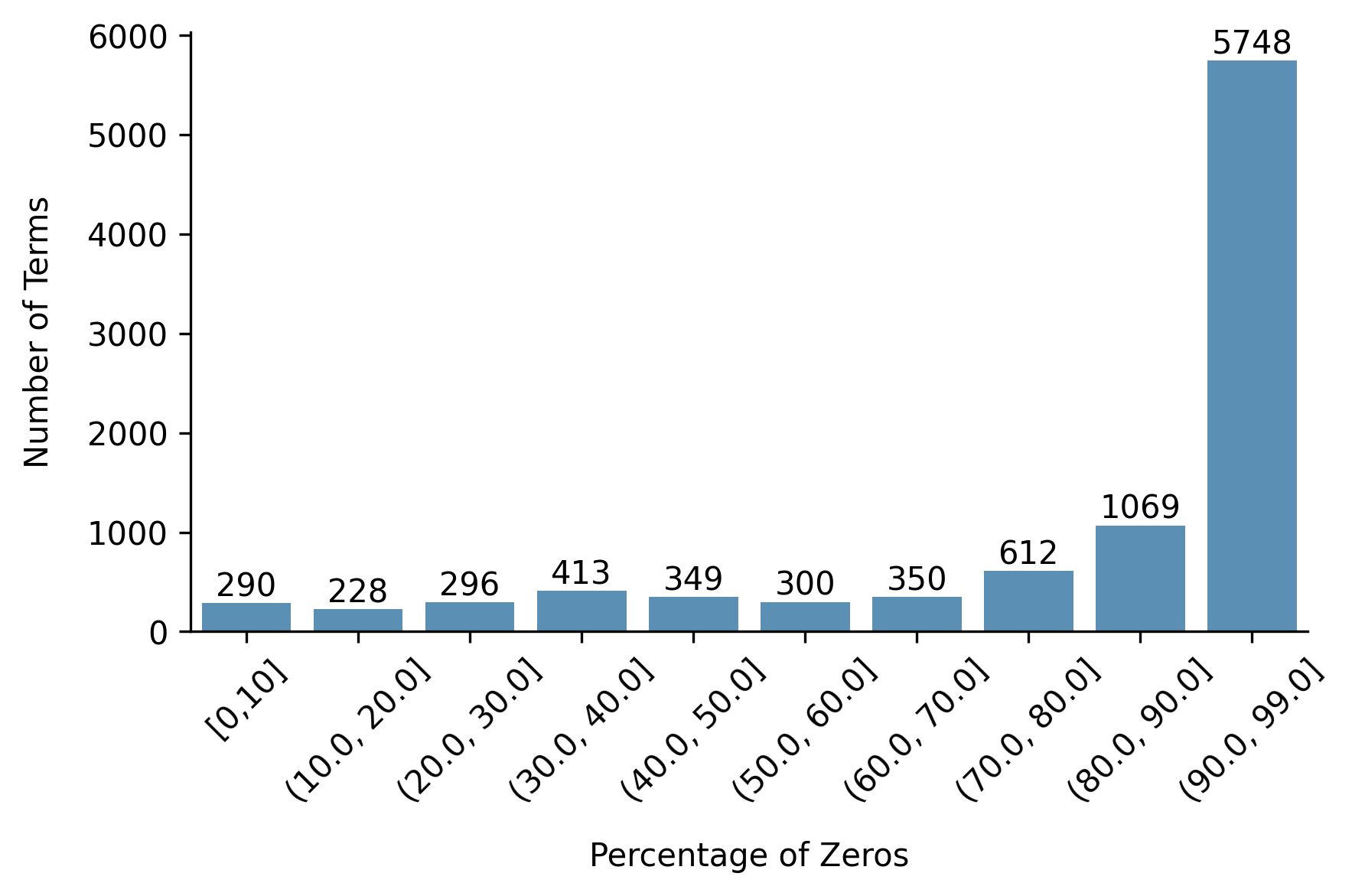}
        \caption{Individual}
        \label{fig:indiv_zeros}
    \end{subfigure}
    \begin{subfigure}[b]{0.3\linewidth}
        \centering
        \includegraphics[width=\linewidth]{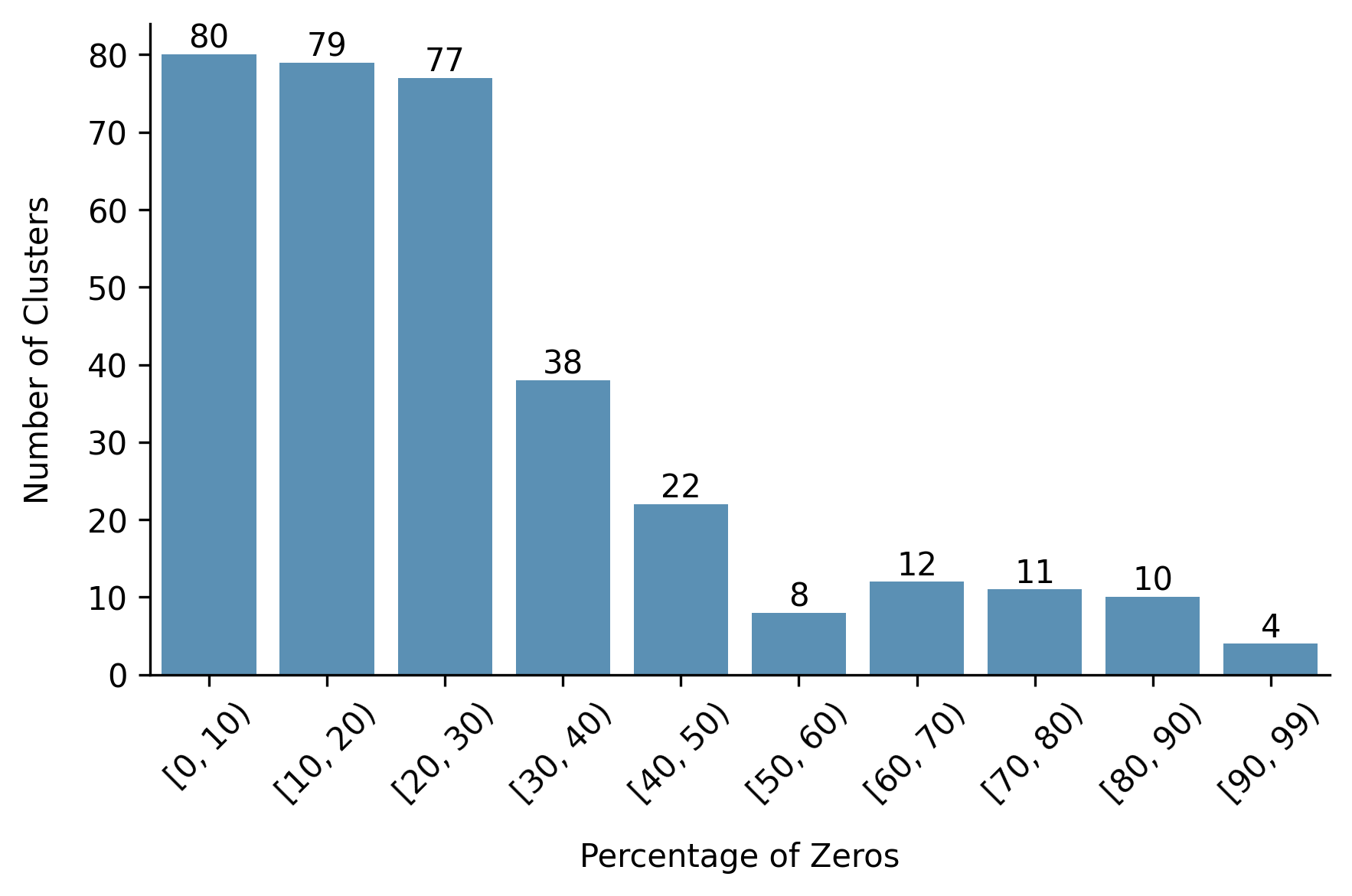}
        \caption{Combined}
        \label{fig:combined_zeros}
    \end{subfigure}
    \caption{Distribution of percentage of zeros in individual and combined search volumes across all locations}
    \label{fig:percentage_zeros}
\end{figure}

Figure \ref{fig:percentage_zeros} analyzes the percentage of missingness before and after clustering for all 28,580 keywords in our dataset. 814 queries with $\leq 30\%$ of zeros were retained in their original form, 8,841 with between 30\% and 99\% zeros were grouped into 341 clusters, and the remainder were discarded. Figure \ref{fig:indiv_zeros} shows that these individual search volumes often contain sparse data with many zeros. After clustering, the percentage of zeros in the aggregated search volumes is notably reduced, as shown in Figure \ref{fig:combined_zeros}. Most clusters fall within the [0–10]\% range of zeros, with very few clusters in the [90–99]\% range, demonstrating a substantial reduction in sparsity. Our method thus transforms sparse data dominated by zeros into denser aggregated data, improving signal quality.

\subsection{Evaluation of Denoising}
\label{sec:noise_results}

Among the 1,155 resulting clusters and individual keywords, smoothing splines are applied to 622 variables requiring denoising. We assess the effectiveness of our smoothing procedure by analyzing noise levels across various downloads retrieved over several months. 

\begin{figure}[H]
    \centering
    \begin{subfigure}{0.4\linewidth}
        \centering
        \includegraphics[width=\linewidth]{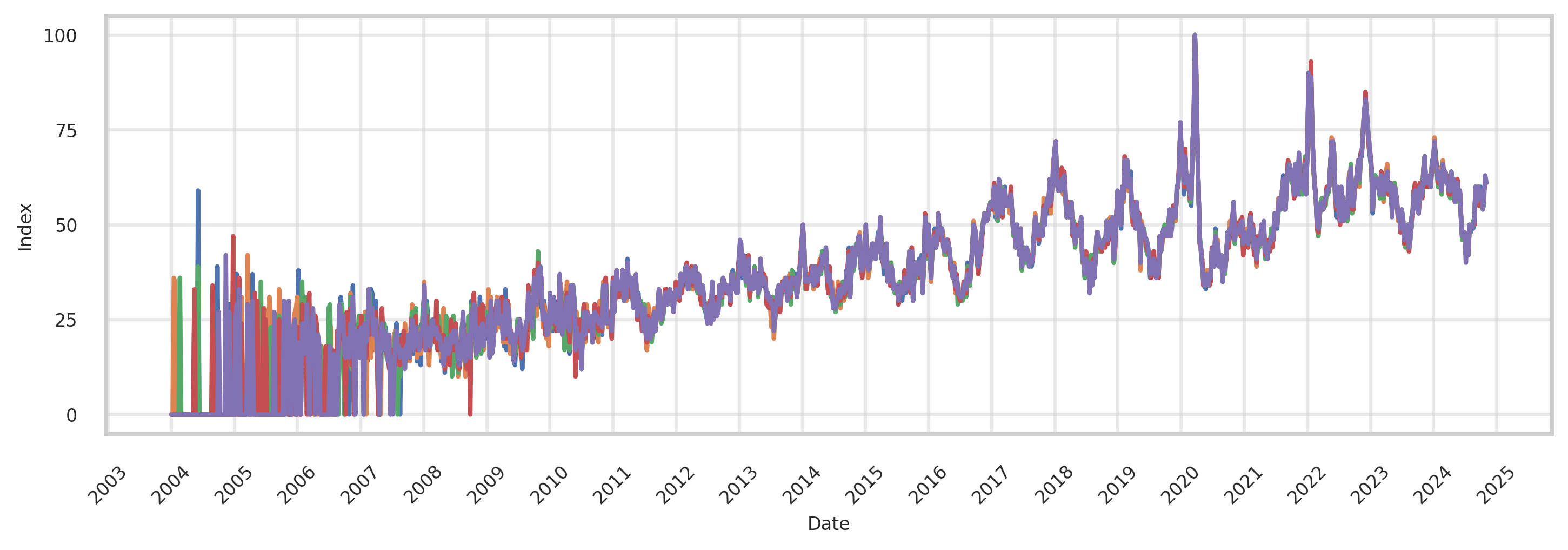}
        \caption{Raw}
        \label{fig:variability_raw}
    \end{subfigure}
    \begin{subfigure}{0.4\linewidth}
        \centering
        \includegraphics[width=\linewidth]{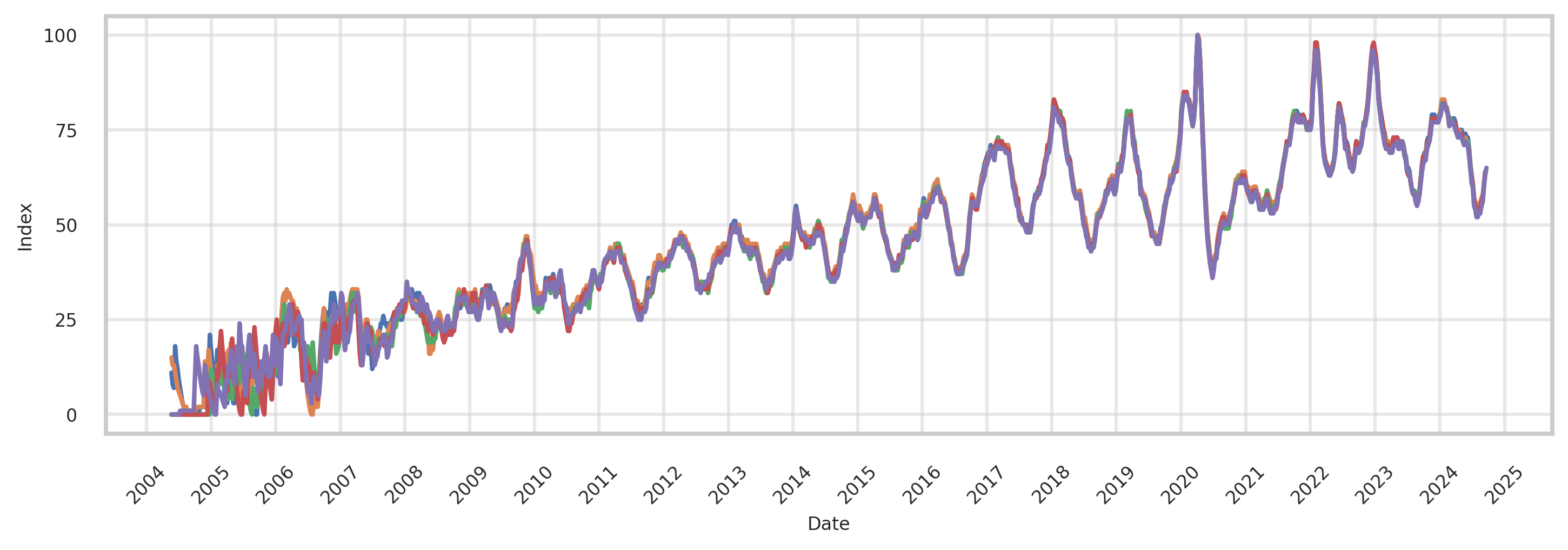}
        \caption{Denoised}
        \label{fig:variability_denoised}
    \end{subfigure}
    \caption{Example comparison of noise levels before and after denoising across five downloads in Alaska}
    \label{fig:variability_comparison}
\end{figure}

\begin{figure}[H]
    \centering
    \begin{subfigure}{0.4\linewidth}
        \centering
        \includegraphics[width=\linewidth]{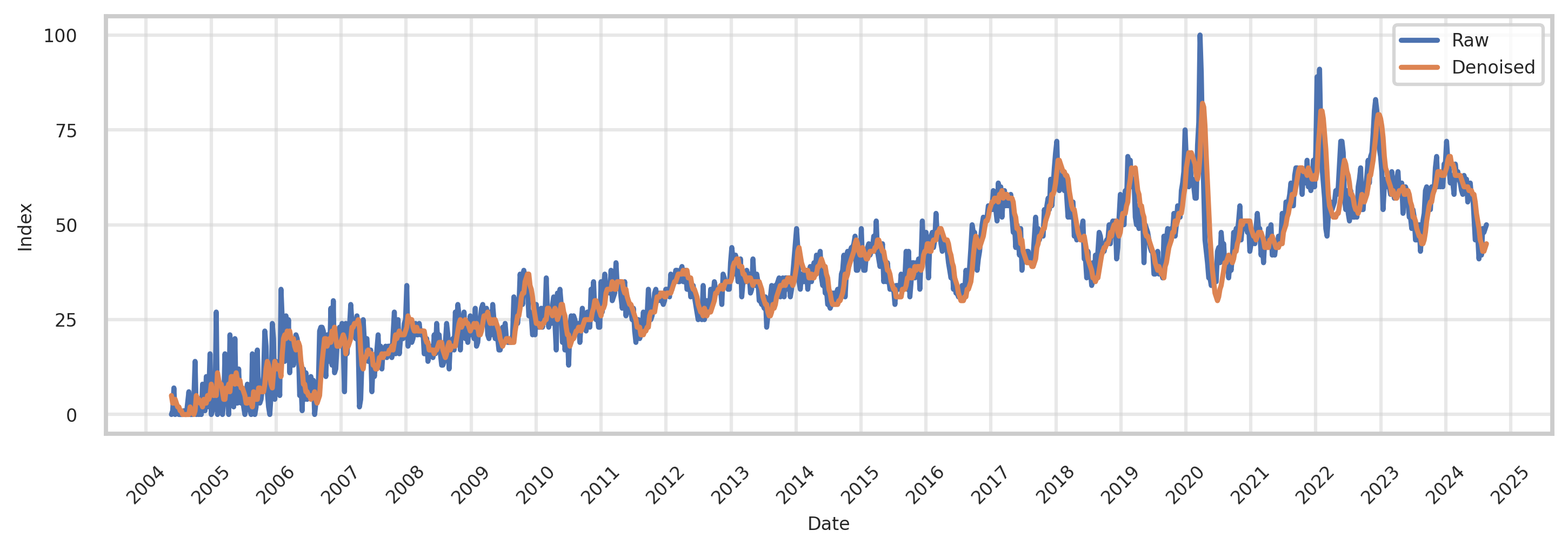}
        \caption{Average}
        \label{fig:avg_raw_denoised}
    \end{subfigure}
    \begin{subfigure}{0.4\linewidth}
        \centering
        \includegraphics[width=\linewidth]{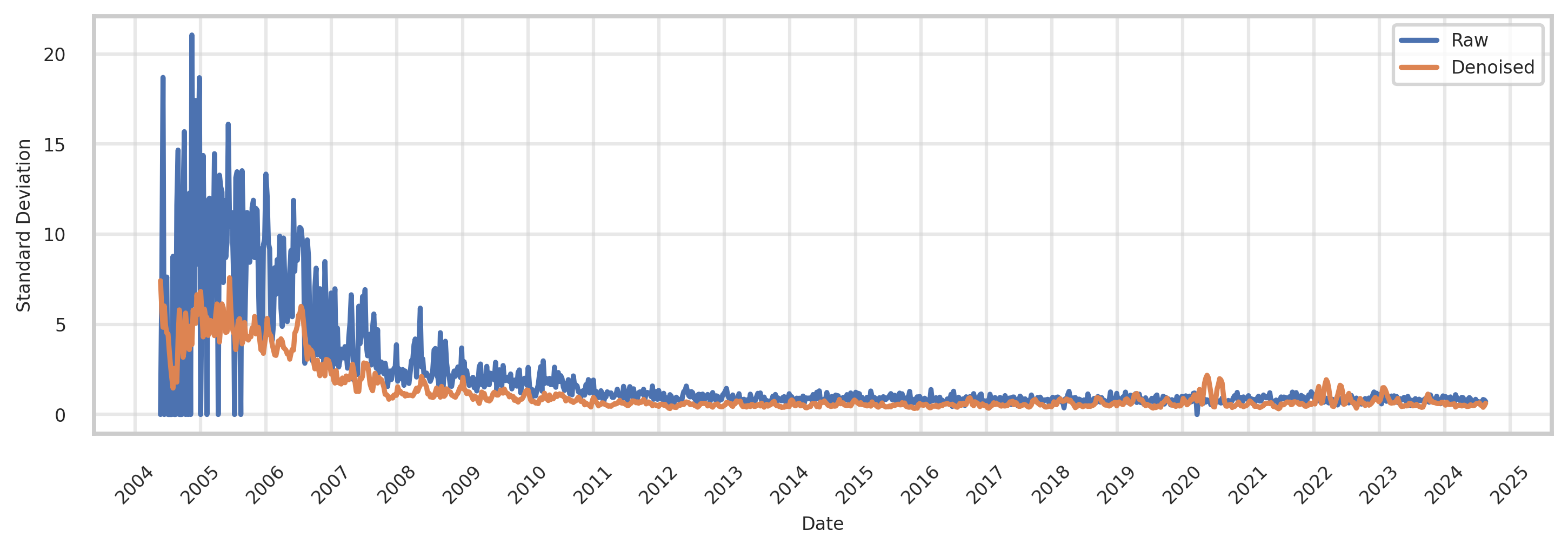}
        \caption{Standard deviation}
        \label{fig:std_raw_denoised}
    \end{subfigure}
    \caption{Descriptive statistics before and after denoising across 27 downloads in Alaska example}
    \label{fig:noise_comparison}
\end{figure}

Figure \ref{fig:variability_comparison} visualizes noise reduction before and after denoising across five repeated downloads for a cluster in Alaska. The raw data in Figure \ref{fig:variability_raw} exhibit substantial noise and sampling variability, while the denoised data in Figure \ref{fig:variability_denoised} display much cleaner signals. Figure \ref{fig:noise_comparison} depicts the averages and standard deviations of the series at each time step across twenty-seven downloads. The averages of raw and denoised data (Figure \ref{fig:avg_raw_denoised}) show similar peaks and troughs, meaning that the smoothing method preserves the overall signal structure and main trends. The lower standard deviation in the denoised data (Figure \ref{fig:std_raw_denoised}) indicates that smoothing reduces sampling variability and noise, leading to a more stable signal.

\begin{figure}[H]
    \centering
    \begin{subfigure}{0.5\linewidth}
        \centering
        \includegraphics[width=\linewidth]{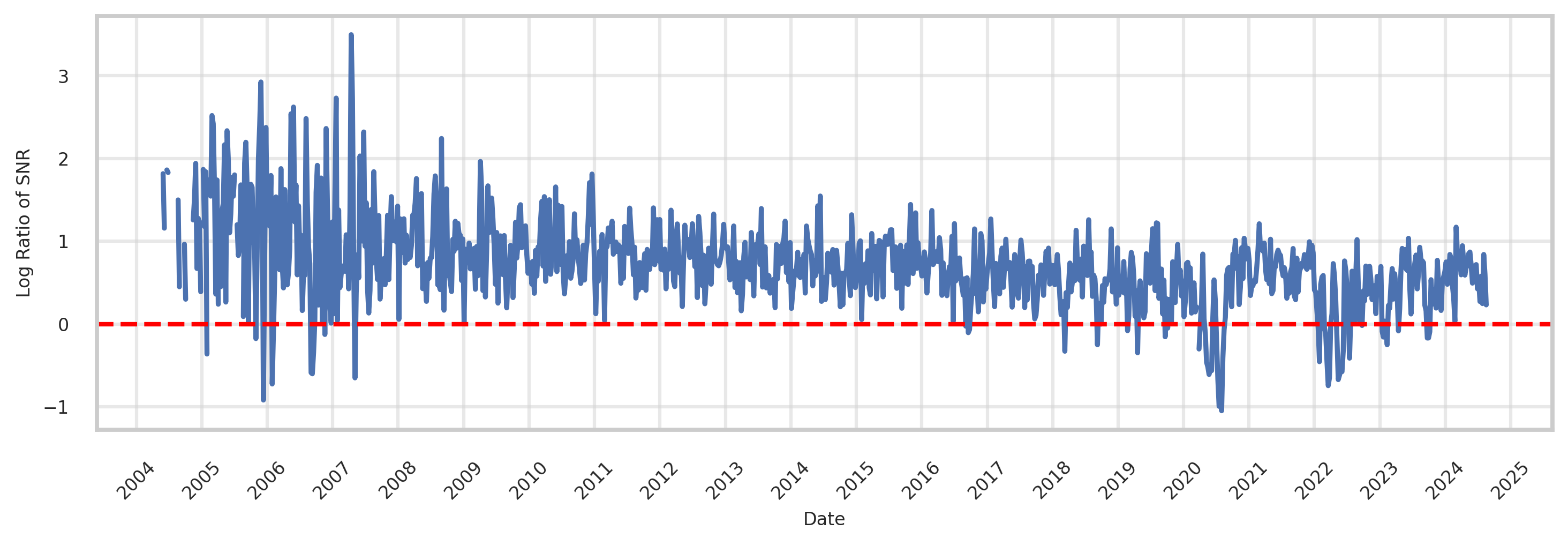}
        \caption{Single keyword in Alaska example}
        \label{fig:ratio_snr_single}
    \end{subfigure}
    \begin{subfigure}{0.3\linewidth}
        \centering
        \includegraphics[width=\linewidth]{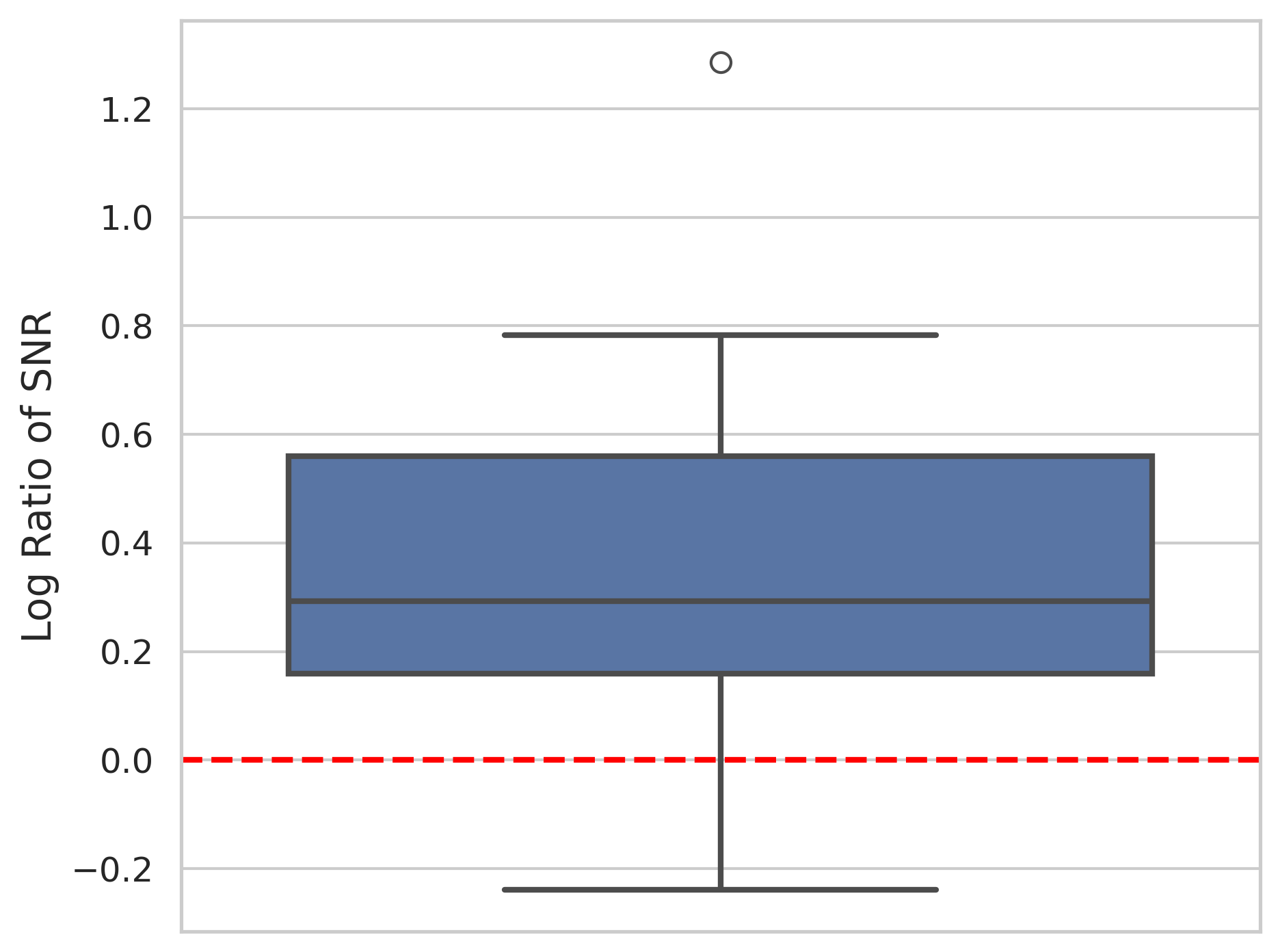}
        \caption{Summary across all keywords and locations}
        \label{fig:ratio_snr_all}
    \end{subfigure}
    \caption{Log ratio of SNR across 27 downloads. Higher is better.}
    \label{fig:snr}
\end{figure}

Effective denoising should minimize short-term fluctuations while preserving underlying patterns. The Signal-to-Noise Ratio (SNR) measures the power of the signal to the power of the noise \citep{palani2022principles}. Since real-world data lacks a clear distinction between signal and noise, \cite{smith1997scientist} define SNR as the ratio of the mean to the standard deviation of the data, $\text{SNR} = \mu / \sigma$, where $\mu$ reflects the overall magnitude of the signal, and $\sigma$ approximates the noise level. A higher SNR indicates a cleaner signal. 

We downloaded repeated instances of the same query at each time point $t$, where each instance is an independent and identically distributed (iid) realization of the underlying signal plus Google-injected noise. Averaging these instances, referred to as signal averaging, reduces the effect of random fluctuations, as noise tends to cancel out when averaged, while the underlying signal is preserved \citep{dempster2001laboratory}.

We estimate the SNR at each time step as $\text{SNR}_t = \frac{\text{mean(repeated)}}{\text{sd(repeated)}} = \hat\mu_t / \hat\sigma_t$, where $\hat\mu_t$ and $\hat\sigma_t$ are the mean and standard deviation of the time series for a query at time $t$ across repeated downloads. This measures how strong the average signal is compared to its variability. For raw and denoised data, we denote their respective SNR as $\text{SNR}^{\text{(raw)}}_t = \hat\mu_t^{\text{(raw)}} / \hat\sigma_t^{\text{(raw)}}$ and $\text{SNR}^{\text{(denoised)}}_t = \hat\mu_t^{\text{(denoised)}} / \hat\sigma_t^{\text{(denoised)}}$. We then compare the two measures by computing the log ratio at each time step $t$, allowing us to track changes in signal quality over time:

\begin{equation}
    \text{log Ratio of SNR} = \log \left( \frac{\hat\mu_t^{\text{(denoised)}}} {\hat\sigma_t^{\text{(denoised)}}} \Big/ \frac{\hat\mu_t^{\text{(raw)}}}{\hat\sigma_t^{\text{(raw)}}} \right).
    \label{eq:log_snr}
\end{equation}

Figure \ref{fig:snr} presents the log ratio of SNR across twenty-seven downloads. Figure \ref{fig:ratio_snr_single} illustrates its temporal evolution for a single time series. Figure \ref{fig:ratio_snr_all} displays the mean log ratio for each keyword. Predominantly positive values indicate that denoised data have higher SNR than raw data, confirming that our denoising procedure enhances signal quality by reducing noise and sampling variability.

\begin{figure}[H]
    \centering
    \includegraphics[width=0.8\linewidth]{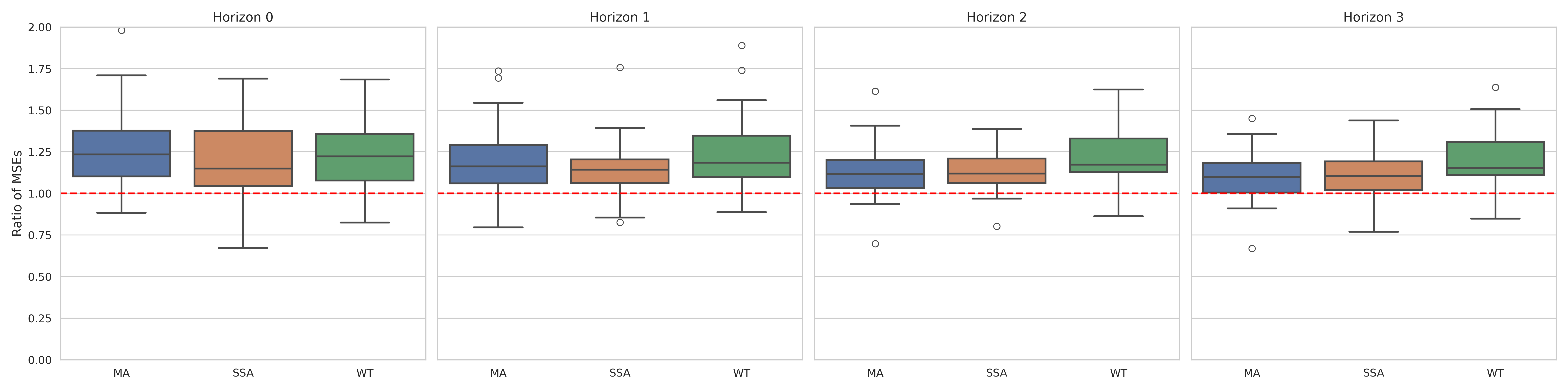}
    \caption{Ratios of MSEs across all locations and horizons for different denoising methods, relative to smoothing splines (benchmark). Values below 1 indicate better performance than smoothing splines.}
    \label{fig:smooth_box}
\end{figure}

We further assess the performance of our proposed denoising technique against existing approaches from the literature, in the context of ARIMAX forecasting. Our framework applies smoothing splines iteratively to smooth time series without introducing look-ahead bias. To ensure a fair comparison, we substitute smoothing splines with alternative filters such as moving averages (MA), singular spectrum analysis (SSA), and wavelet transform (WT), while keeping the workflow unchanged. Implementation details are provided in Section \ref{app:denoising}. Figure \ref{fig:smooth_box} shows the ratios of MSEs from forecasts using each alternative method relative to those using smoothing splines across all keywords, locations and horizons. The boxplots are mostly above 1, indicating that smoothing splines deliver the most accurate forecasts.

\subsection{Evaluation of Detrending}
\label{sec:trend_results}

Detrending is applied after denoising to keywords that exhibit a trend based on the ADF test. To assess its effectiveness in removing deterministic trends, we use the regression coefficient of determination ($R^2$), which quantifies how well the estimated trend explains the variability of the data. This metric is computed on the entire time series between the estimated trend and the data. A lower $R^2$ after detrending indicates successful trend removal.

\begin{figure}[H]
    \centering
    \includegraphics[width=0.5\linewidth]{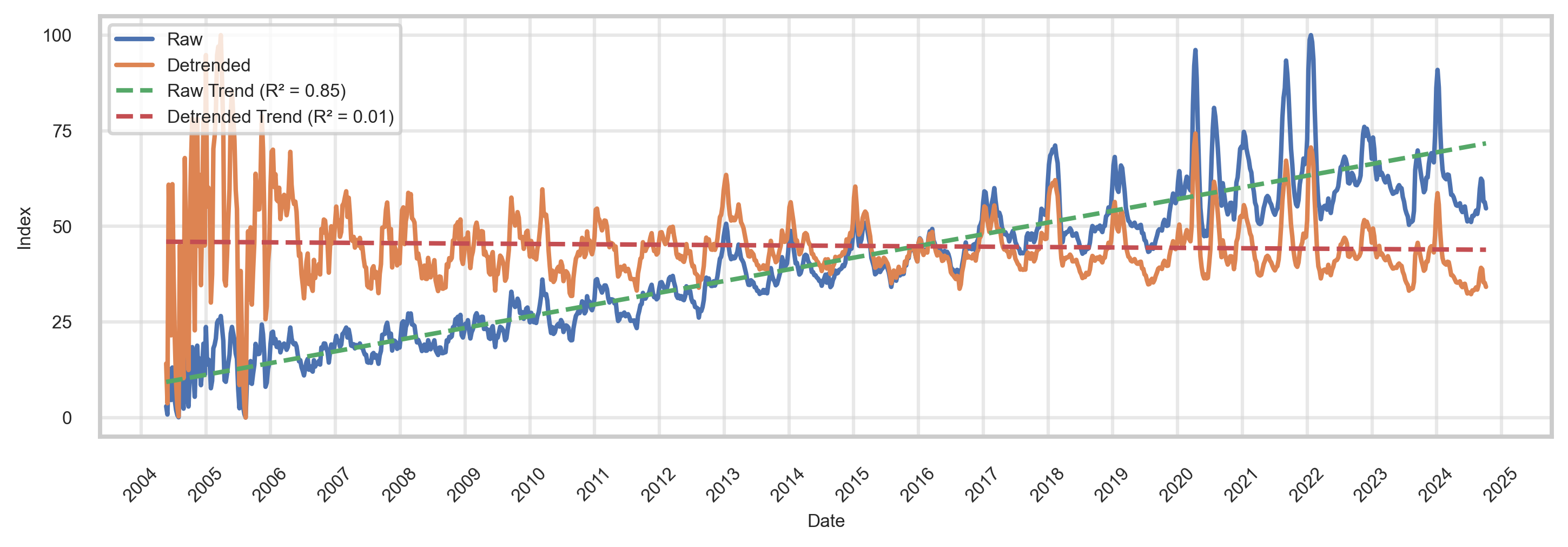}
    \caption{Example of estimated deterministic trend in raw and detrended data in Alabama}
    \label{fig:raw_detrend}
\end{figure}

Figure \ref{fig:raw_detrend} displays the time series of a cluster in Alabama before and after detrending, along with their respective fitted trends. Before detrending, $R^2$ equals 0.85, meaning the trend accounts for 85\% of the variability. After detrending, it drops to 0.01, confirming that the trend no longer dominates the data.

\begin{figure}[H]
    \centering
    \includegraphics[width=0.3\linewidth]{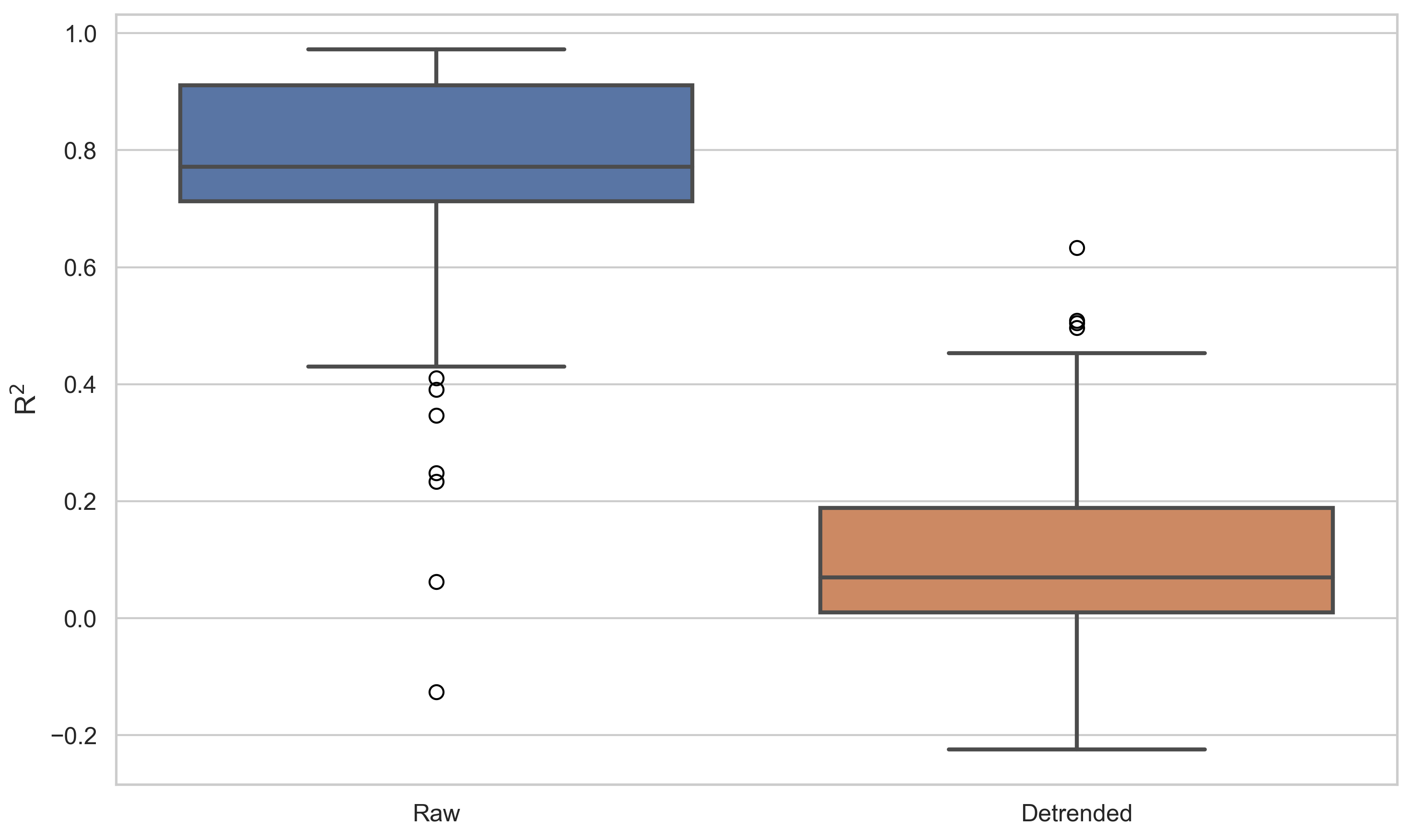}
    \caption{Summary of $R^2$ for deterministic trends before and after detrending across all locations}
    \label{fig:r2_trends}
\end{figure}

Figure \ref{fig:r2_trends} summarizes the distribution of $R^2$ across all keywords. Fitted trends in the raw data exhibit high $R^2$ values (mean $= 0.77$), confirming that they were well estimated. After detrending, $R^2$ values decrease significantly (mean $ = 0.11$), demonstrating the effective removal of deterministic trends. There are a few outliers with low $R^2$ values before detrending and high values afterward, suggesting occasional misestimation, as stationarity is detected based on the training set and may not fully capture time series patterns. 

\section{Application to Influenza Prediction}
\label{sec:end-to-end}

For a real-world validation of our preprocessing methodology, we apply the pipeline to the task of real-time influenza forecasting several weeks ahead using both statistical and machine learning models. We retrospectively evaluate state-level forecasts of U.S. influenza hospitalizations (including DC) across two seasons, from October 17, 2022, to April 27, 2024, with training data extending up to October 2022. The application combines Google Trends data of flu-related search terms with hospitalization counts from the U.S. Centers for Disease Control and Prevention (CDC), demonstrating the practical utility of the proposed methodology.

\subsection{Identifying Flu-related Terms from Google Trends}
\label{sec:flu_terms}

In this study, we analyze weekly Google Trends' ``interest over time'' data for the United States at the state level since 2004. Our initial set of 161 flu-related terms and topics is based on the work of \citet{yang2015accurate}, identified through Google Correlate in 2009 and 2010, but the service was discontinued in 2019. Therefore, to expand our pool of keywords, we retrieve queries related to ``flu''. For long-term seasonal analysis, ``top'' queries are preferred over ``rising'' ones, as they exhibit more consistent search patterns in the given time frame. We choose the ``Cold and Flu'' category to limit results to the area of infectious diseases, and select the entire period from 2004 onward to exclude event-specific queries (e.g., ``flu 2024''). Keywords are tailored to each state by setting the corresponding location, which also restricts results to English (or the predominant local language), as opposed to collecting worldwide data. In summary, category, time period, and location serve as filters to refine the dataset. This approach yields time series of weekly search volumes for approximately 500 terms per state, covering symptoms, treatments, medicine, or types of flu.

\subsection{Influenza Hospitalizations Data from CDC}
\label{sec:cdc}

The target variable in our case study is influenza-related hospitalizations from the US Centers for Disease Control and Prevention. Weekly admission counts have been collected at the state level in the United States since February 2, 2022 \citep{mathis2024evaluation} by the National Healthcare Safety Network (NHSN) \citep{nhsn}. This data is publicly available through CDC's GitHub repository \citep{flusightgithub}. To extend the training period for modeling, we follow the methodology of \citet{meyer2025prospective} to augment the dataset using ILINet data \citep{cdcili} and FluSurv-Net data \citep{cdchosp} with transfer learning. This augmentation provides continuous time series of hospitalization data starting from 2012. Influenza-related hospitalization data is released weekly with a one-week delay, while Google Trends data is available in real time, offering a timely informational advantage when used as exogenous variables in forecasting.

\subsection{Modeling}

For linear statistical approaches, we consider ARIMAX \citep{box2015time, tiao1981modeling}, seasonal ARIMAX \citep{box2015time, tiao1981modeling}, and ARGO (AutoRegression with GOogle search data) \citep{yang2015accurate}, all of which are widely used and well-established for influenza forecasting \citep{ray2018prediction, kandula2019near, reich2019collaborative}. For nonlinear machine learning methods, we apply the tree-based methods LightGBM (Light Gradient Boosting Machine) \citep{ke2017lightgbm} and AdaBoost (adaptive boosting) \citep{freund1997decision}, both of which have demonstrated strong performance in influenza prediction \citep{meyer2025prospective, ray2025flusion, santillana2015combining}. Implementation details are provided in Section \ref{app:model_specs}

We augment each model with Google Trends data to evaluate the contribution of our preprocessing approach to forecasting performance. Each method is applied separately for every location to produce retrospective out-of-sample nowcasts and 1–3 week-ahead forecasts, simulating real-time predictions. This setup allows us to compare forecast performance across various scenarios: using preprocessed data, topics only, non-preprocessed data, and omitting exogenous variables as baselines.

The pool of keywords is obtained through Google Trends' ``related queries'', which may include terms lacking semantic relevance that cannot be manually excluded. The raw non-preprocessed dataset of individual queries is very large, therefore, we reduce it by either using topics only or by setting a zero-filter to remove keywords with too many zeros.  The preprocessed dataset is much smaller, as queries are combined. To identify meaningful predictors for both raw and preprocessed data, we apply a correlation-based filter, selecting variables most correlated with the target for each location while limiting the total number of predictors to avoid overfitting. To prevent multicollinearity, we drop one variable from any pair with high correlation. Correlations are computed exclusively on the training set to prevent look-ahead bias and preserve the integrity of the real-time forecasting framework. 

\subsection{Forecasting Results and Statistical Significance}

Forecasting performance is assessed using the Mean Squared Error (MSE) between the ground truth and predicted values, $\text{MSE} = \frac{1}{T} \sum_{t=1}^T (y_t - \hat{y}_t)^2$, over the test period $t=1,...,T$. We then compute the Relative Efficiency (RE) as the ratio of the MSE for each model with exogenous variables to the MSE of the corresponding model without exogenous variables, which serves as the benchmark. An RE below 1 indicates improved performance over the baseline.

To evaluate statistical significance, we apply the Wilcoxon Signed-Rank Test \citep{wilcoxon1945} to the out-performance of each model. Specifically, we conduct a one-sided paired test with the null hypothesis that the median of paired differences in MSE is equal to 0 against the alternative that the median is less than 0. This nonparametric test is robust to outliers and well-suited for our 51 paired MSEs with heteroskedasticity (one per state) for each model comparison. Table \ref{tab:summary_re} reports the median RE, interquartile range, and p-values aggregated across all locations, horizons, models, and sets of exogenous variables. P-values below 0.05 are indicated as $*$ for comparisons against the baseline without exogenous information, $\dagger$ for preprocessed (clustering, denoising, and detrending) versus non-preprocessed data, and $\ddagger$ for preprocessed versus topic-only data.

\begin{table}[H]
\centering
\resizebox{\textwidth}{!}{%
\begin{tabular}{ll *{5}{l c}}
\toprule
{Horizon} & {Method} & \multicolumn{2}{c}{ARIMAX} & \multicolumn{2}{c}{SARIMAX} & \multicolumn{2}{c}{ARGO} & \multicolumn{2}{c}{LightGBM} & \multicolumn{2}{c}{AdaBoost} \\
\cmidrule(lr){3-4} \cmidrule(lr){5-6} \cmidrule(lr){7-8} \cmidrule(lr){9-10} \cmidrule(lr){11-12}
&& Median & (Q1, Q3) & Median & (Q1, Q3) & Median & (Q1, Q3) & Median & (Q1, Q3) & Median & (Q1, Q3) \\
\midrule
\multirow{5}{*}{0}
    & Detrending & \textbf{0.83}$^{*\dagger\ddagger}$ & (0.75, 0.94) & \textbf{1.01}$^{\dagger\ddagger}$ & (0.85, 1.17) & 0.83$^{*\dagger}$ & (0.69, 0.94) & 0.85$^{*\dagger\ddagger}$ & (0.76, 0.95) & 0.81$^{*}$ & (0.65, 0.98) \\
    & Denoising  & 0.85$^{*\dagger\ddagger}$ & (0.75, 0.91) & 1.04$^{\dagger\ddagger}$ & (0.84, 1.18) & 0.80$^{*\dagger}$ & (0.67, 0.92) & \textbf{0.83}$^{*\dagger\ddagger}$ & (0.76, 0.90) & \textbf{0.80}$^{*\dagger}$ & (0.66, 0.95) \\
    & Clustering & 0.92$^{*\dagger\ddagger}$ & (0.80, 0.97) & \textbf{1.01}$^{\dagger\ddagger}$ & (0.84, 1.23) & \textbf{0.76}$^{*\dagger}$ & (0.55, 0.88) & 0.88$^{*\ddagger}$ & (0.79, 0.97) & 0.83$^{*\dagger}$ & (0.68, 0.94) \\
    & Topics     & 1.32 & (1.20, 1.63) & 1.21 & (1.03, 1.57) & 0.78$^{*}$ & (0.60, 0.96) & 0.93$^{*}$ & (0.84, 1.00) & 0.82$^{*}$ & (0.71, 0.97) \\
    & Non-preprocessed & 1.65 & (1.50, 1.87) & 1.93 & (1.37, 2.81) & 0.94 & (0.82, 1.13) & 0.89$^{*}$ & (0.81, 0.98) & 0.83$^{*}$ & (0.73, 1.03) \\
\midrule
\multirow{5}{*}{1}
    & Detrending & \textbf{0.80}$^{*\dagger\ddagger}$ & (0.69, 0.86) & 0.81$^{*\dagger\ddagger}$ & (0.68, 0.96) & 0.72$^{*\dagger\ddagger}$ & (0.59, 0.86) & \textbf{0.71}$^{*\dagger\ddagger}$ & (0.64, 0.86) & 0.62$^{*\dagger}$ & (0.36, 0.74) \\
    & Denoising  & 0.82$^{*\dagger\ddagger}$ & (0.70, 0.87) & 0.82$^{*\dagger\ddagger}$ & (0.70, 0.93) & 0.71$^{*\dagger\ddagger}$ & (0.59, 0.86) & 0.75$^{*\dagger\ddagger}$ & (0.60, 0.83) & \textbf{0.58}$^{*\dagger}$ & (0.34, 0.76) \\
    & Clustering & 0.82$^{*\dagger\ddagger}$ & (0.71, 0.92) & \textbf{0.80}$^{*\dagger\ddagger}$ & (0.67, 0.98) & \textbf{0.66}$^{*\dagger\ddagger}$ & (0.50, 0.81) & 0.80$^{*}$ & (0.70, 0.94) & 0.71$^{*}$ & (0.49, 0.83) \\
    & Topics     & 1.17 & (1.06, 1.31) & 0.95 & (0.76, 1.12) & 0.92$^{*}$ & (0.75, 1.06) & 0.80$^{*}$ & (0.73, 1.02) & 0.59$^{*}$ & (0.43, 0.86) \\
    & Non-preprocessed & 1.27 & (1.16, 1.43) & 1.05 & (0.82, 1.44) & 1.05 & (0.89, 1.27) & 0.85$^{*}$ & (0.71, 0.97) & 0.75$^{*}$ & (0.55, 0.91) \\
\midrule
\multirow{5}{*}{2}
    & Detrending & \textbf{0.83}$^{*\dagger\ddagger}$ & (0.71, 0.88) & \textbf{0.72}$^{*\dagger\ddagger}$ & (0.61, 0.90) & 0.73$^{*\dagger\ddagger}$ & (0.56, 0.84) & \textbf{0.75}$^{*\dagger}$ & (0.65, 0.88) & \textbf{0.61}$^{*\dagger}$ & (0.44, 0.82) \\
    & Denoising  & 0.84$^{*\dagger\ddagger}$ & (0.73, 0.88) & 0.75$^{*\dagger\ddagger}$ & (0.60, 0.88) & 0.71$^{*\dagger\ddagger}$ & (0.55, 0.86) & 0.76$^{*\dagger\ddagger}$ & (0.67, 0.83) & 0.64$^{*\dagger}$ & (0.49, 0.83) \\
    & Clustering & 0.84$^{*\dagger\ddagger}$ & (0.72, 0.92) & 0.76$^{*\dagger\ddagger}$ & (0.59, 0.85) & \textbf{0.63}$^{*\dagger\ddagger}$ & (0.48, 0.82) & 0.85$^{*}$ & (0.73, 0.97) & 0.75$^{*}$ & (0.58, 0.93) \\
    & Topics     & 1.11 & (1.01, 1.23) & 0.90 & (0.69, 1.12) & 1.12 & (1.02, 1.25) & 0.83$^{*}$ & (0.75, 0.98) & 0.68$^{*}$ & (0.45, 0.86) \\
    & Non-preprocessed & 1.18 & (1.04, 1.37) & 0.93 & (0.68, 1.20) & 1.12 & (1.06, 1.41) & 0.86$^{*}$ & (0.78, 1.03) & 0.77$^{*}$ & (0.60, 0.97) \\
\midrule
\multirow{5}{*}{3}
    & Detrending & \textbf{0.82}$^{*\dagger\ddagger}$ & (0.72, 0.90) & \textbf{0.72}$^{*\dagger\ddagger}$ & (0.54, 0.89) & 0.73$^{*\dagger\ddagger}$ & (0.54, 0.87) & \textbf{0.78}$^{*\dagger}$ & (0.66, 0.99) & \textbf{0.64}$^{*\dagger}$ & (0.51, 0.79) \\
    & Denoising  & 0.84$^{*\dagger\ddagger}$ & (0.73, 0.91) & \textbf{0.72}$^{*\dagger\ddagger}$ & (0.59, 0.86) & 0.74$^{*\dagger\ddagger}$ & (0.53, 0.87) & 0.81$^{*}$ & (0.69, 1.04) & 0.72$^{*\dagger}$ & (0.55, 0.86) \\
    & Clustering & 0.86$^{*\dagger\ddagger}$ & (0.73, 0.94) & 0.78$^{*\dagger\ddagger}$ & (0.61, 0.89) & \textbf{0.67}$^{*\dagger\ddagger}$ & (0.47, 0.80) & 0.89$^{*}$ & (0.74, 1.04) & 0.78$^{*\dagger}$ & (0.60, 0.90) \\
    & Topics     & 1.08 & (0.99, 1.22) & 0.87 & (0.68, 1.13) & 1.07 & (1.00, 1.18) & 0.88$^{*}$ & (0.72, 1.02) & 0.75$^{*}$ & (0.55, 0.90) \\
    & Non-preprocessed & 1.15 & (1.02, 1.33) & 0.88$^{*}$ & (0.67, 1.11) & 1.15 & (0.98, 1.27) & 0.89$^{*}$ & (0.73, 1.03) & 0.81$^{*}$ & (0.67, 0.98) \\
\bottomrule
\end{tabular}
} 
\vspace{5pt}
\caption{Median and interquartile range (Q1, Q3) of relative efficiency across all locations, horizons, models, and sets of exogenous variables. Values below 1 indicate better performance than the benchmark model without exogenous variables. Boldfaced medians correspond to the best performance relative to the baseline. p-values from the Wilcoxon Signed-Rank Test  are reported as $*: p < 0.05$ for models with exogenous data against baselines; $\dagger: p < 0.05$ for preprocessed versus non-preprocessed data; and $\ddagger: p < 0.05$ for preprocessed versus topic-only data.}
\label{tab:summary_re}
\end{table}

Across all models and horizons, incorporating at least one preprocessing step consistently improves forecasts relative to omitting exogenous variables or using raw or topic-only data. The p-values for preprocessed versus baseline comparisons fall well below 0.05, strongly supporting rejection of the null hypothesis and confirming that preprocessing significantly reduces forecast errors. Similarly, comparisons against non-preprocessed and topic-only data show that at least one step of preprocessing outperforms raw data in most models and horizons. In contrast, using raw data degrades forecasting performance in statistical models relative to the baseline. Although the optimal level of preprocessing varies by model, these results highlight the overall benefit of applying our preprocessing pipeline to Google Trends data. Figure \ref{fig:all_box} provides boxplots illustrating variation across locations. 

\subsection{Comparison of Different Preprocessing Steps in Time-Varying Forecast Performance}

We analyze forecasting performance separately during peak and off-season periods to assess how each preprocessing step contributes across different phases of influenza activity. Figure \ref{fig:time_variation_all} illustrates that forecast differences between clustered, denoised, and detrended data are most pronounced during flu peaks (December 2022 and January 2024). Table \ref{tab:peak_off_mses} reports the average RE accross all locations of detrending and denoising relative to clustering alone for each forecast horizon and model. Overall, clustering by itself performs best during peak periods, as indicated by RE values above 1 for most models, whereas denoising and detrending improve accuracy during the off-season. These results suggest that denoising and detrending may overly smooth short-term fluctuations that are informative during peak outbreaks, but provide clear benefits when forecasting outside the peak season.

\begin{table}[H]
\centering
\resizebox{0.8\textwidth}{!}{%
\begin{tabular}{l l l c c c c c}
\toprule
{Horizon} & {Season} & {Method} & ARIMAX & SARIMAX & ARGO & LightGBM & AdaBoost \\ 
\midrule
\multirow{6}{*}{0} 
    & \multirow{3}{*}{Peak} 
        & Detrending & 0.85 & 1.01 & 0.89 & 1.05 & 1.01 \\ 
    & & Denoising  & 0.89 & 0.97 & 0.89 & 1.05 & 1.01 \\
\cmidrule(lr){3-8}
    & \multirow{3}{*}{Off} 
        & Detrending & 0.61 & 0.86 & 0.82 & 0.94 & 1.00 \\ 
    & & Denoising  & 0.57 & 0.80 & 0.74 & 0.94 & 1.00 \\ 
\midrule
\multirow{6}{*}{1} 
    & \multirow{3}{*}{Peak} 
        & Detrending & 1.11 & 1.10 & 0.89 & 1.00 & 1.05 \\ 
    & & Denoising  & 1.08 & 0.98 & 0.89 & 1.00 & 1.05 \\ 
\cmidrule(lr){3-8}
    & \multirow{3}{*}{Off} 
        & Detrending & 0.78 & 0.81 & 0.95 & 1.03 & 0.93 \\ 
    & & Denoising  & 0.78 & 0.81 & 0.90 & 1.03 & 0.93 \\ 
\midrule
\multirow{6}{*}{2} 
    & \multirow{3}{*}{Peak} 
        & Detrending & 1.07 & 1.08 & 0.94 & 1.00 & 1.01 \\ 
    & & Denoising  & 1.06 & 0.98 & 0.95 & 1.01 & 1.01 \\ 
\cmidrule(lr){3-8}
    & \multirow{3}{*}{Off} 
        & Detrending & 0.89 & 0.84 & 1.32 & 1.06 & 0.99 \\ 
    & & Denoising  & 0.92 & 0.81 & 1.16 & 1.05 & 0.99 \\ 
\midrule
\multirow{6}{*}{3} 
    & \multirow{3}{*}{Peak} 
        & Detrending & 1.01 & 0.99 & 0.97 & 0.9 & 1.00\\ 
    & & Denoising  & 1.01 & 0.93 & 1.04 & 0.9 & 1.00\\ 
\cmidrule(lr){3-8}
    & \multirow{3}{*}{Off} 
        & Detrending & 1.02 & 0.91 & 1.70 & 1.04 & 0.88\\ 
    & & Denoising  & 1.04 & 0.87 & 1.50 & 1.02 & 0.88\\ 
\bottomrule
\end{tabular}
} 
\vspace{5pt}
\caption{Average relative efficiency between detrending/denoising and clustering across all locations, horizons, and models during peak (December–January) and off (February–November) periods over two influenza seasons. Values below 1 indicate better performance than clustering.}
\label{tab:peak_off_mses}
\end{table}

To examine how forecast performance evolves over time, we apply the Fluctuation Test of \citet{giacomini2010forecast} pairwise to each model's errors using clustered, denoised, and detrended data. Unlike simple comparisons of average errors, this test assesses whether the relative performance of two methods remains stable over time. Applied to average errors across all locations for each model, the two-sided test does not reject the null hypothesis for any horizon or any pair of exogenous datasets at the confidence level 0.05. Therefore, we find no evidence that the relative performance of clustered, denoised, or detrended predictors varies over time when aggregated across locations. In other words, the three preprocessing steps perform equally well throughout the evaluation period. Figures \ref{fig:fluctuation_h0}-\ref{fig:fluctuation_h3} display the average sample paths of the Fluctuation Test across locations for each pair of preprocessing methods, model, and forecast horizon.

\section{Discussion and Conclusion}
\label{sec:discussion}

In this paper, we investigate the data generation mechanism of Google Trends and propose a comprehensive statistical preprocessing methodology to maximize its utility for forecasting. Our approach integrates hierarchical clustering to address missing values and sampling variability, dynamic smoothing splines to reduce noise, and detrending to separate short-term signals from long-term trends. Each preprocessing step is independently evaluated to validate its effectiveness in improving search data quality. We further demonstrate the practical benefits of our approach through an application to influenza hospitalization forecasting, from nowcasts up to three-week-ahead predictions, showing improved accuracy when incorporating preprocessed search data into a statistical or machine learning model. 

While our method has proven effective, it is not without limitations. First, keyword selection relied on Google Trends’ ``related queries'' feature, which occasionally included terms unrelated to flu. Rather than manually excluding these terms, we mitigated their impact through clustering, which reduces the influence of a single irrelevant term on the aggregated search volumes, and by applying a correlation filter to exclude terms with low relevance to the target variable. Second, the clustering process occasionally grouped semantically unrelated terms, but these clusters remained meaningful in terms of their time series patterns, ensuring their utility in prediction. Third, the quality of Google Trends data varies significantly across locations, with regions of lower internet penetration or smaller populations producing inherently lower quality signals that preprocessing cannot fully address. For example, California benefits from a large volume of searches, resulting in higher-quality data, while Alaska’s smaller search base produces more limited signals. This heterogeneity affects the number of meaningful clusters that can be identified in each location. In our study, we extracted a total of 56 individual terms, topics, and clusters for California, compared with only 5 for Alaska, illustrating how data sparsity potentially limits predictive ability in less populous regions. 

We evaluate the effectiveness of our preprocessing techniques using a range of models, from linear statistical models, such as ARIMAX, SARIMAX, and ARGO, to nonlinear machine learning models, including LightGBM and AdaBoost. We choose interpretable models to highlight the direct contributions of Google Trends data. As each model has different specifications and levels of complexity, the appropriate level of preprocessing varies. ARIMAX and SARIMAX, which use only a few past target values, rely heavily on exogenous inputs for predictive information. For these models, denoising prevents overfitting to noise, and detrending aligns exogenous variables with the differenced target, both of which are crucial for accurate forecasts. In comparison, ARGO incorporates up to 52 weeks of past target values, which capture long-term trends, seasonal structure, and autocorrelation. In this model, exogenous variables primarily provide short-term, real-time spikes, so denoising and detrending may inadvertently remove informative fluctuations. LightGBM is a robust machine learning model that reduces the number of features through its Exclusive Feature Bundling (EFB) technique, which bundles mutually exclusive features together so that variables that contribute little to reducing prediction error are ignored \citep{ke2017lightgbm}. AdaBoost performs internal feature selection as an integral part of its procedure, since at each iteration the algorithm selects the feature that yields the best split under the current weighting \citep{hastie2017elements}. As a result, these models can outperform simple baselines even with raw input data, though preprocessing further enhances performance. For instance, LightGBM, as a gradient boosting decision tree model, cannot extrapolate trends beyond the range of training values \citep{joseph2022modern}, making detrending of exogenous predictors beneficial. AdaBoost is sensitive to noisy data, as it amplifies errors of mispredicted observations \citep{leistner2009robustness}, and therefore benefits from denoising. Overall, clustering is the most fundamental step of our methodology, as it overcomes privacy thresholds and reduces data sparsity by generating aggregated Google Trends search volumes. Denoising and detrending may be beneficial depending on the model. Across both statistical and machine learning approaches, at least one level of preprocessing appropriately enhances the predictive value of Google Trends data.

Future research could investigate the generalizability of our preprocessing methodology under different conditions, expanding on this study’s focus on enhancing Google Trends data quality for flu hospitalization prediction in the United States. A natural extension would be to apply our approach to other locations, where variations in search behavior, vocabulary, historical trends, and data availability may impact performance. Differences in internet penetration and regional interest could result in varying levels of missing data or noisier patterns, requiring adjustments in preprocessing. Similarly, our methodology could be tested on other infectious diseases or even non-health-related forecasting tasks, as long as search data remains a relevant predictor of the target. Identifying meaningful search terms may be more challenging in domains where prior knowledge is limited, unlike influenza, where symptoms and seasonal patterns are well understood. Moreover, other search-based data sources beyond Google Trends, such as those from search engines, social media platforms, or AI-driven tools, could provide additional signals for forecasting. Understanding how these alternative data sources differ in terms of missing values, sampling variability, noise, and trends, and whether they require similar preprocessing strategies, would help establish the robustness and stability of our methodology across a wider range of applications.



\bibstyle{plain}
\bibliography{paperbibliography}

\setcounter{section}{0}

\renewcommand{\thesection}{S\arabic{section}}
\setcounter{figure}{0}
\renewcommand{\thefigure}{S\arabic{figure}}
\setcounter{table}{0}
\renewcommand{\thetable}{S\arabic{table}}
\newpage

\section*{Supporting Information}

\section{Google Trends}
\label{app:google_trends}

\begin{figure}[h]
    \centering
    \includegraphics[width=1\linewidth]{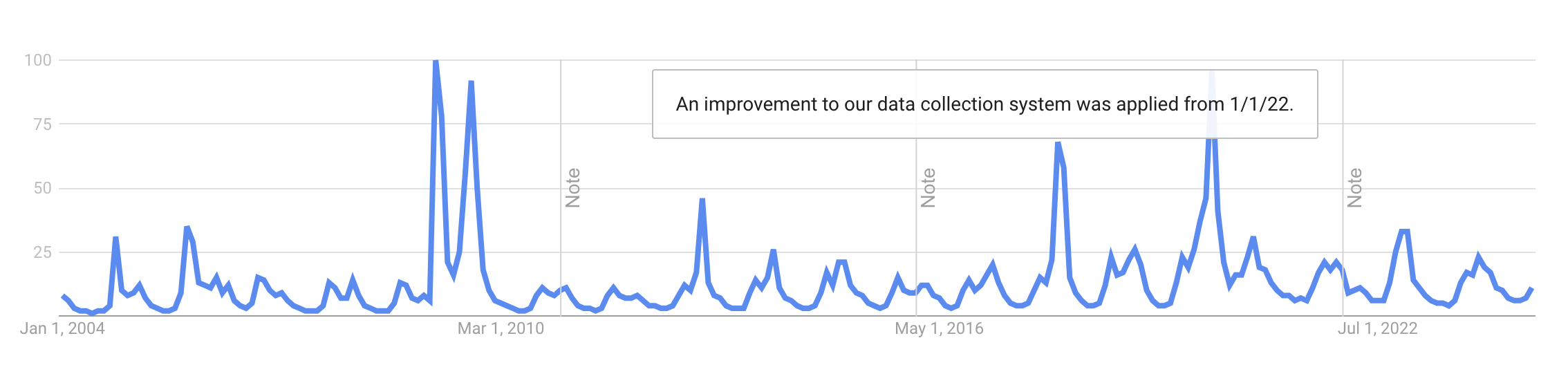}
    \caption{Google Trends website (accessed 2024-09-13) showing the ``interest over time'' for the search term ``flu'' from 2004 to 2024. The website includes annotations indicating data collection updates.}
    \label{fig:gt_notes}
\end{figure}

\begin{table}[h!]
    \centering
    \begin{tabular}{lcc}
        \toprule
        Date & Percentage \\ 
        \midrule
        2024-02-11 & 64\% \\ 
        2024-02-25 & 84\% \\ 
        2024-03-10 & 84\% \\ 
        2024-03-24 & 84\% \\ 
        2024-04-07 & 86\% \\ 
        2024-04-21 & 87\% \\ 
        \bottomrule
    \end{tabular}
    \caption{Overall percentage of zeros across multiple downloads of Google Trends search volumes for the same 161 keywords across 52 locations.}
    \label{tab:zeros_updates}
\end{table}

\section{Forecasting Results}
\label{app:results}

\begin{figure}[H]
    \centering
    \includegraphics[width=1\linewidth]{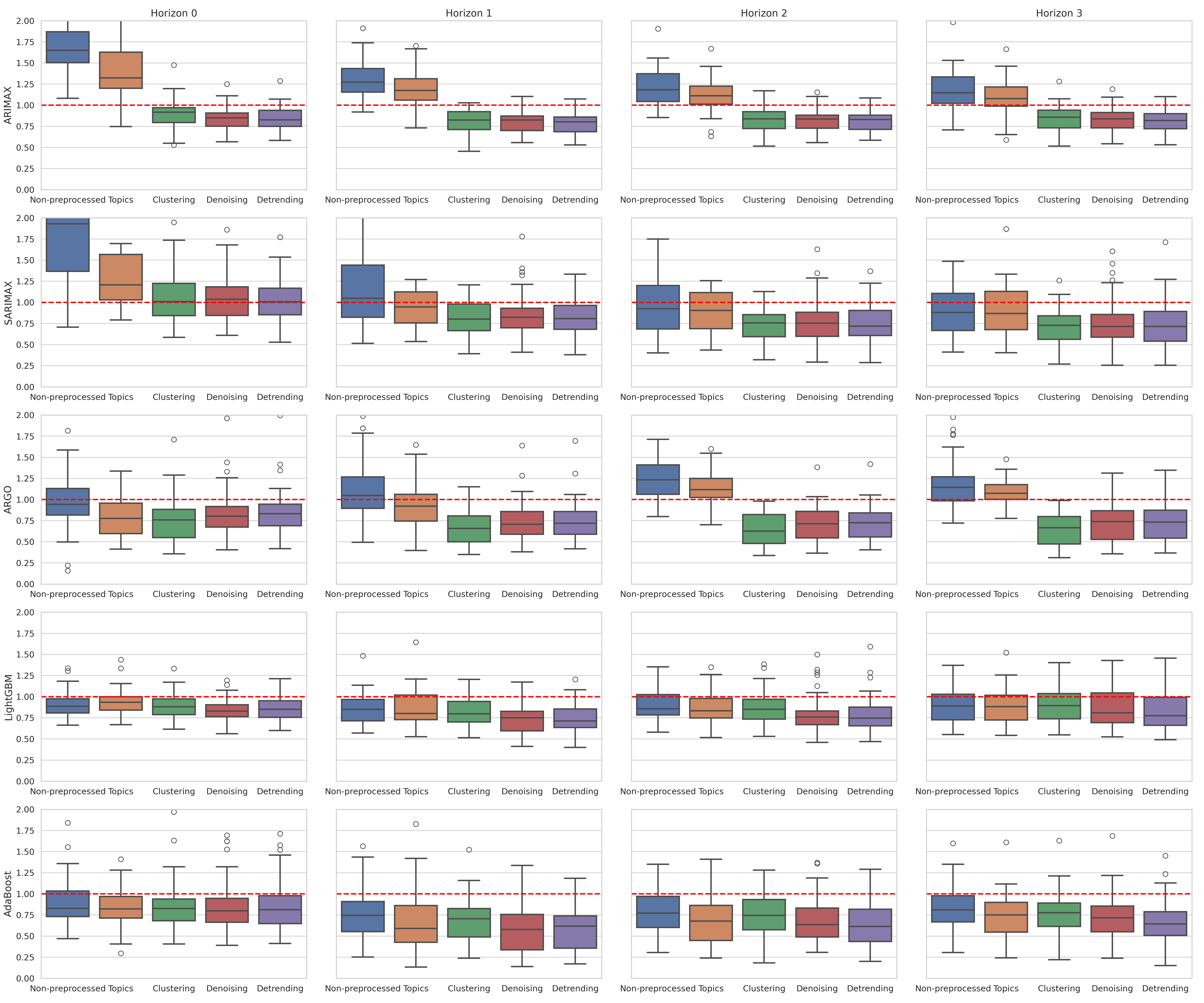}
    \caption{Ratios MSEs across all locations and horizons for different datasets, relative to the model without exogenous variables (baseline). Values below 1 indicate better performance than the baseline.}
    \label{fig:all_box}
\end{figure}

\section{Comparison of Different Preprocessing Steps in Time-Varying Forecast Performance}
\label{app:time_variation}

\begin{figure}[H]
    \centering
    \begin{subfigure}[t]{0.3\linewidth}
        \includegraphics[width=\linewidth]{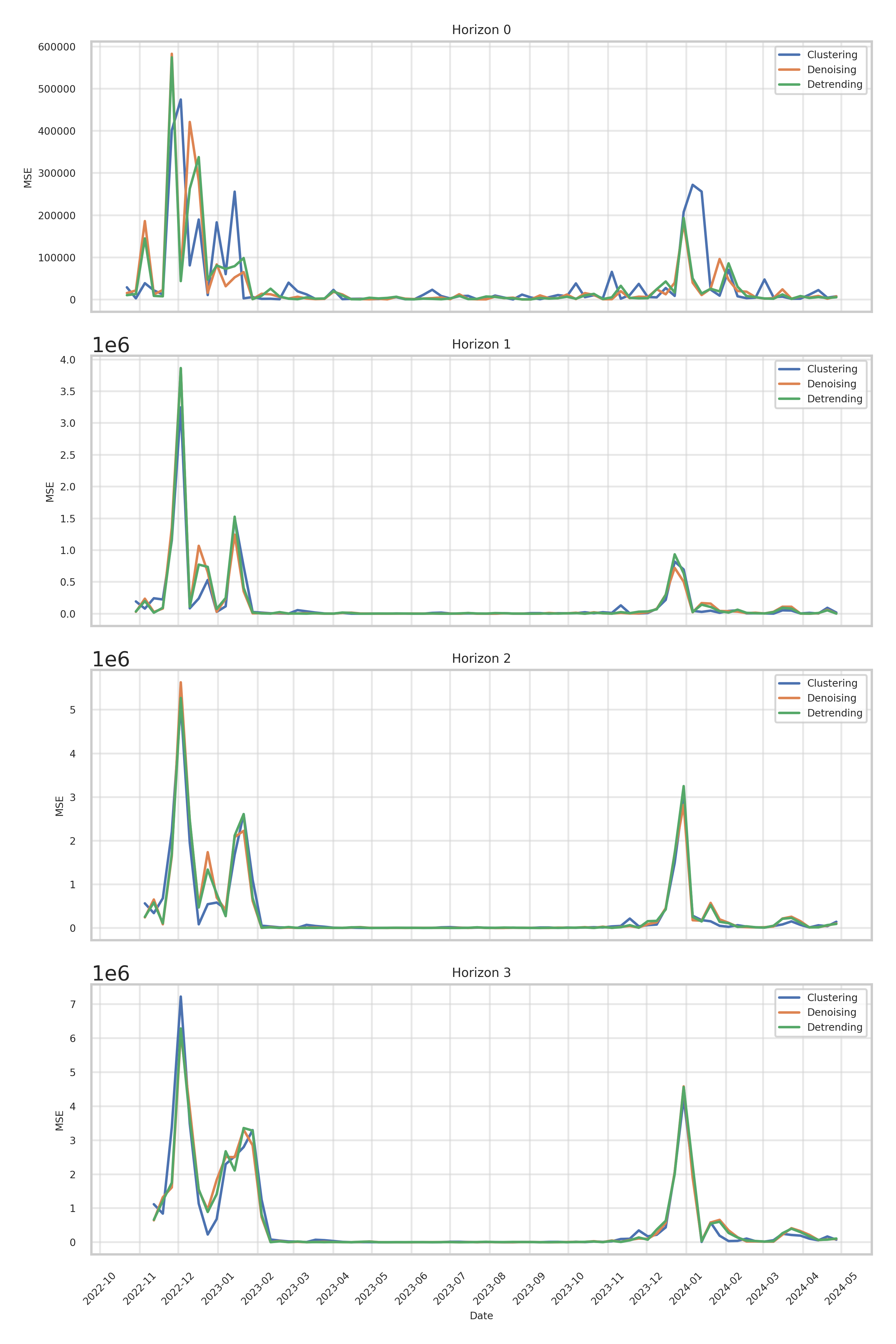}
        \caption{ARIMAX}
        \label{fig:arimax}
    \end{subfigure}
    \begin{subfigure}[t]{0.3\linewidth}
        \includegraphics[width=\linewidth]{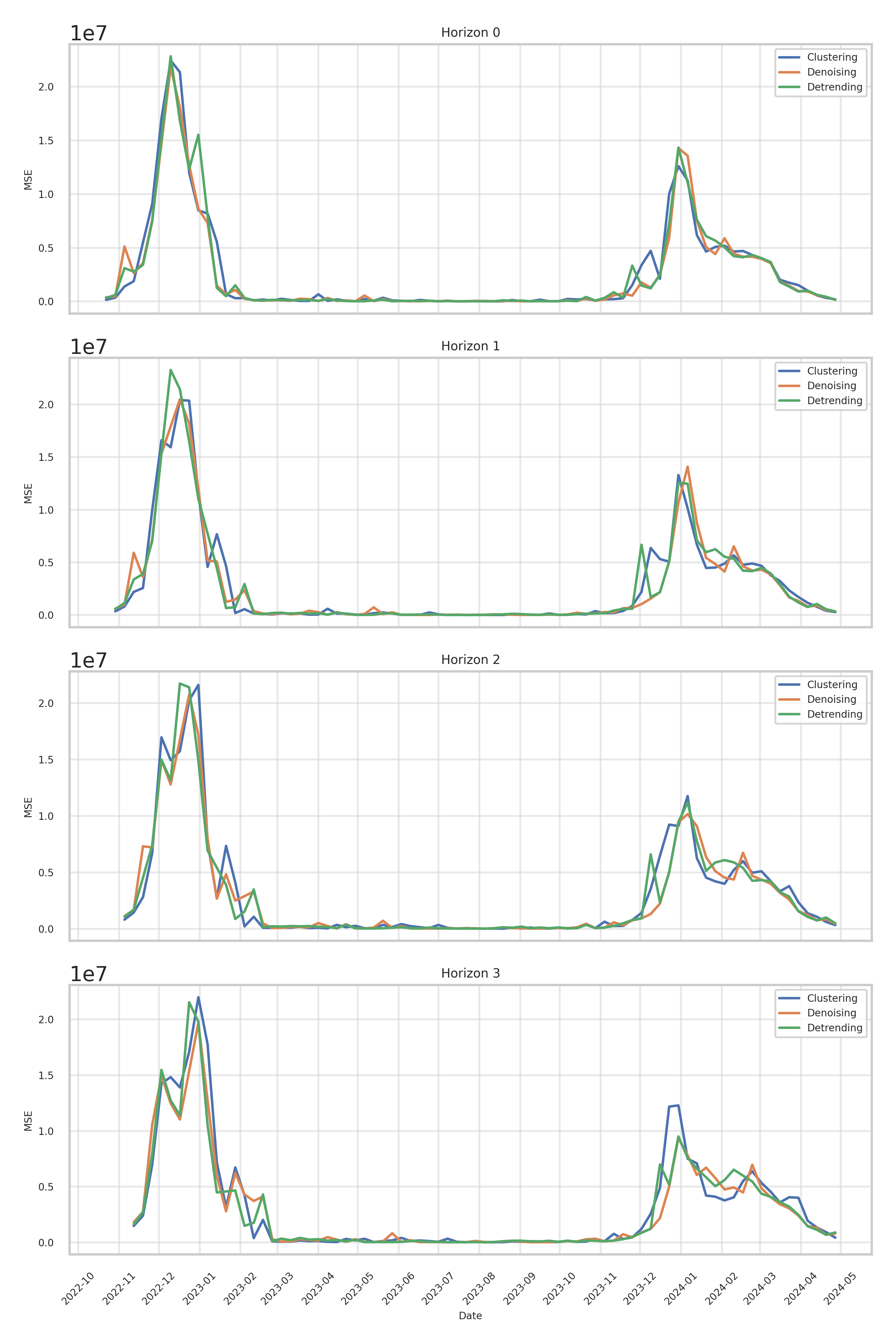}
        \caption{SARIMAX}
        \label{fig:sarimax}
    \end{subfigure}
    \begin{subfigure}[t]{0.3\linewidth}
        \includegraphics[width=\linewidth]{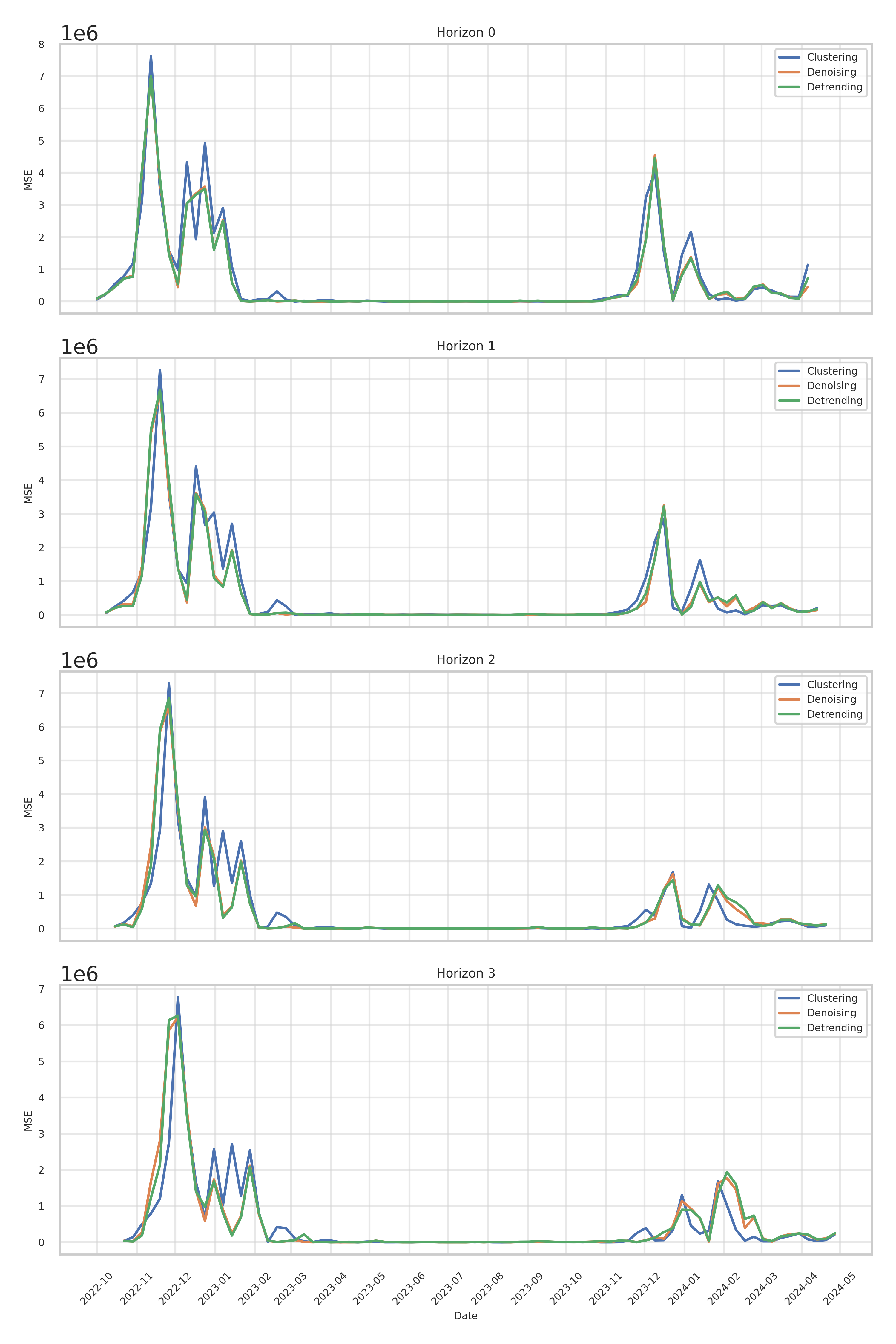}
        \caption{ARGO}
        \label{fig:argo}
    \end{subfigure}
    \begin{subfigure}[t]{0.3\linewidth}
        \includegraphics[width=\linewidth]{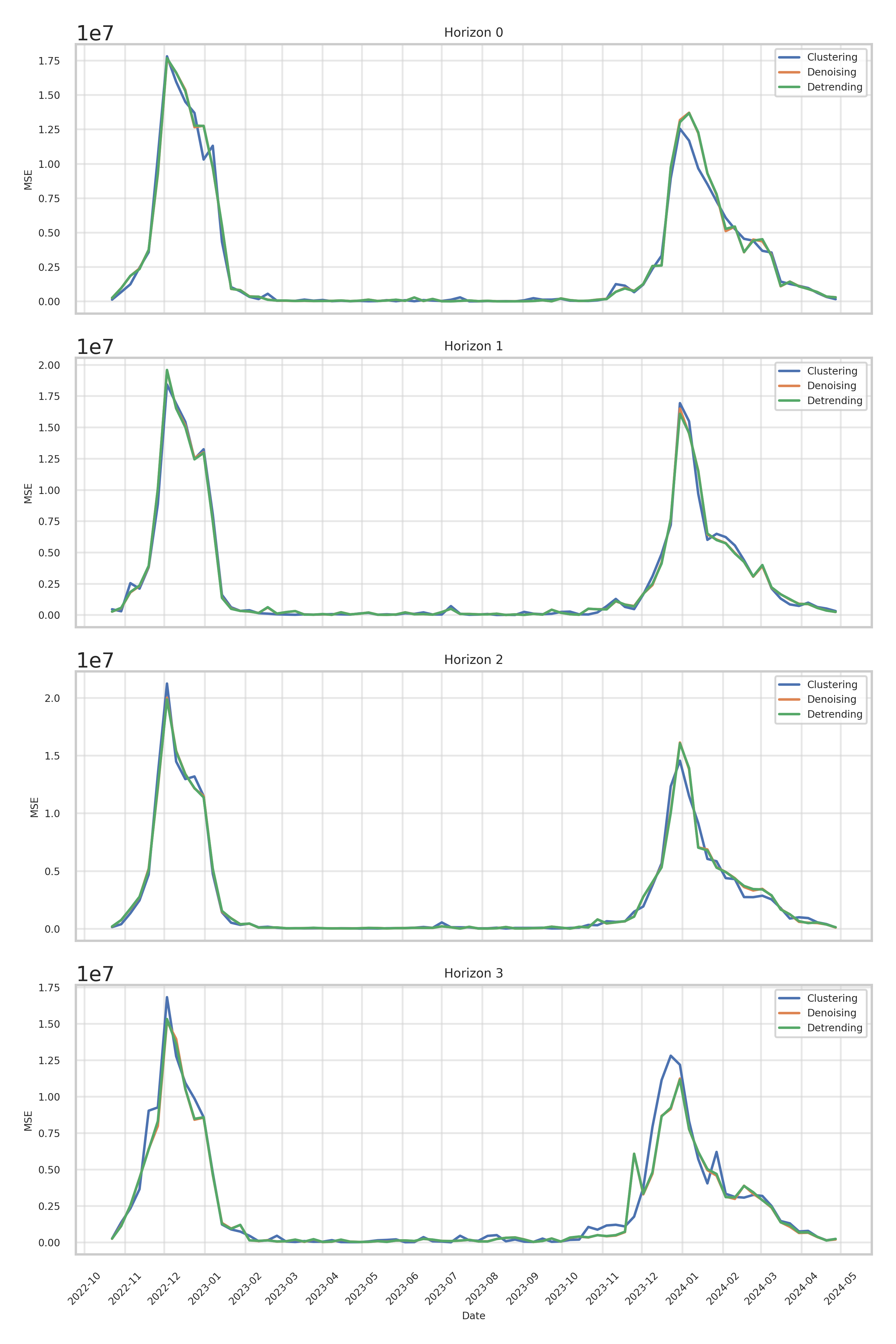}
        \caption{LightGBM}
        \label{fig:lgbm}
    \end{subfigure}
    \begin{subfigure}[t]{0.3\linewidth}
        \includegraphics[width=\linewidth]{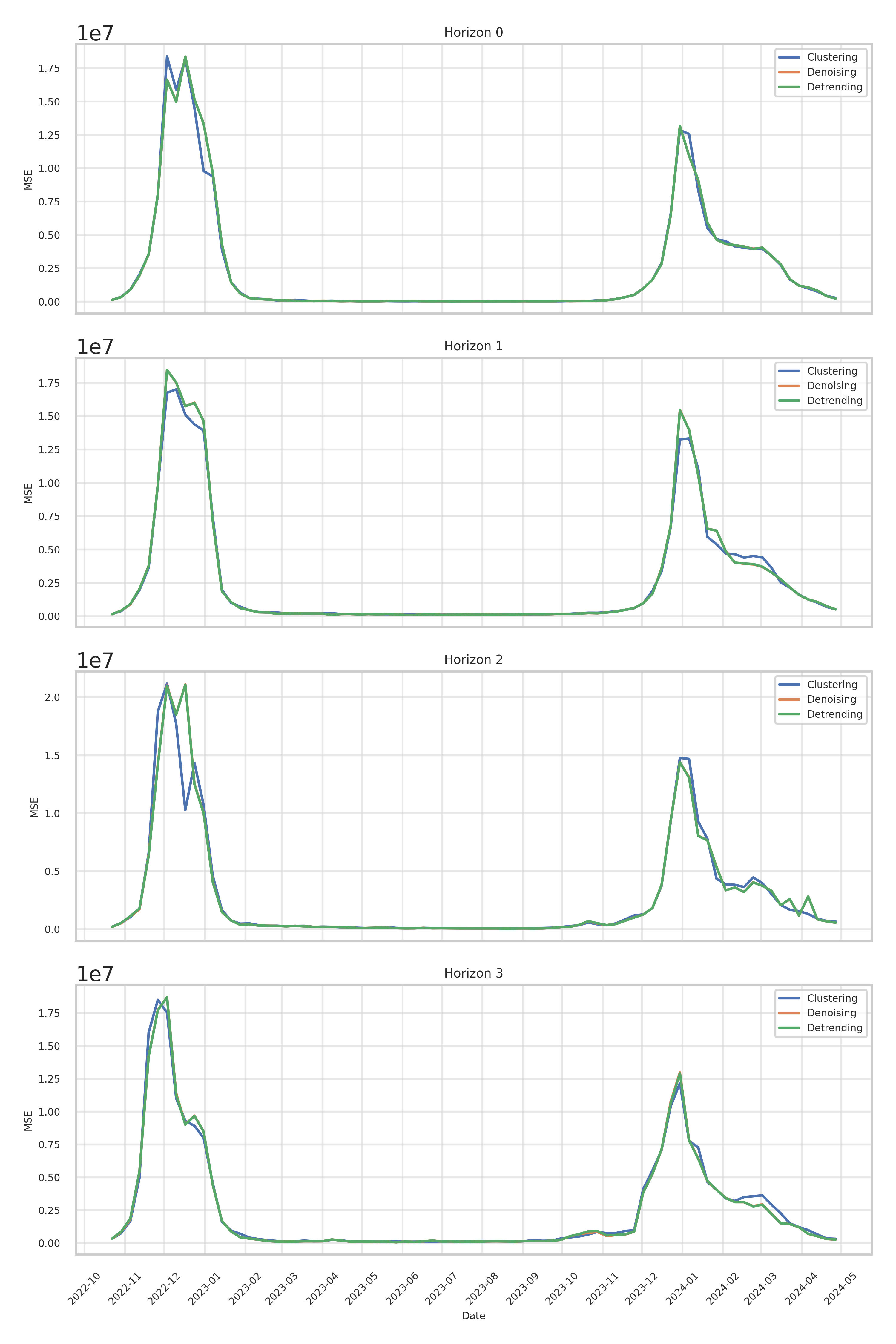}
        \caption{AdaBoost}
        \label{fig:adaboost}
    \end{subfigure}
    \caption{Average weekly MSEs for each model across locations from October 2022 to May 2024 for clustered, denoised, and detrended data.  Lower values indicate better performance. Differences are mostly observable during peaks.}
    \label{fig:time_variation_all}
\end{figure}

\begin{figure}[H]
    \centering
    \includegraphics[width=1\linewidth]{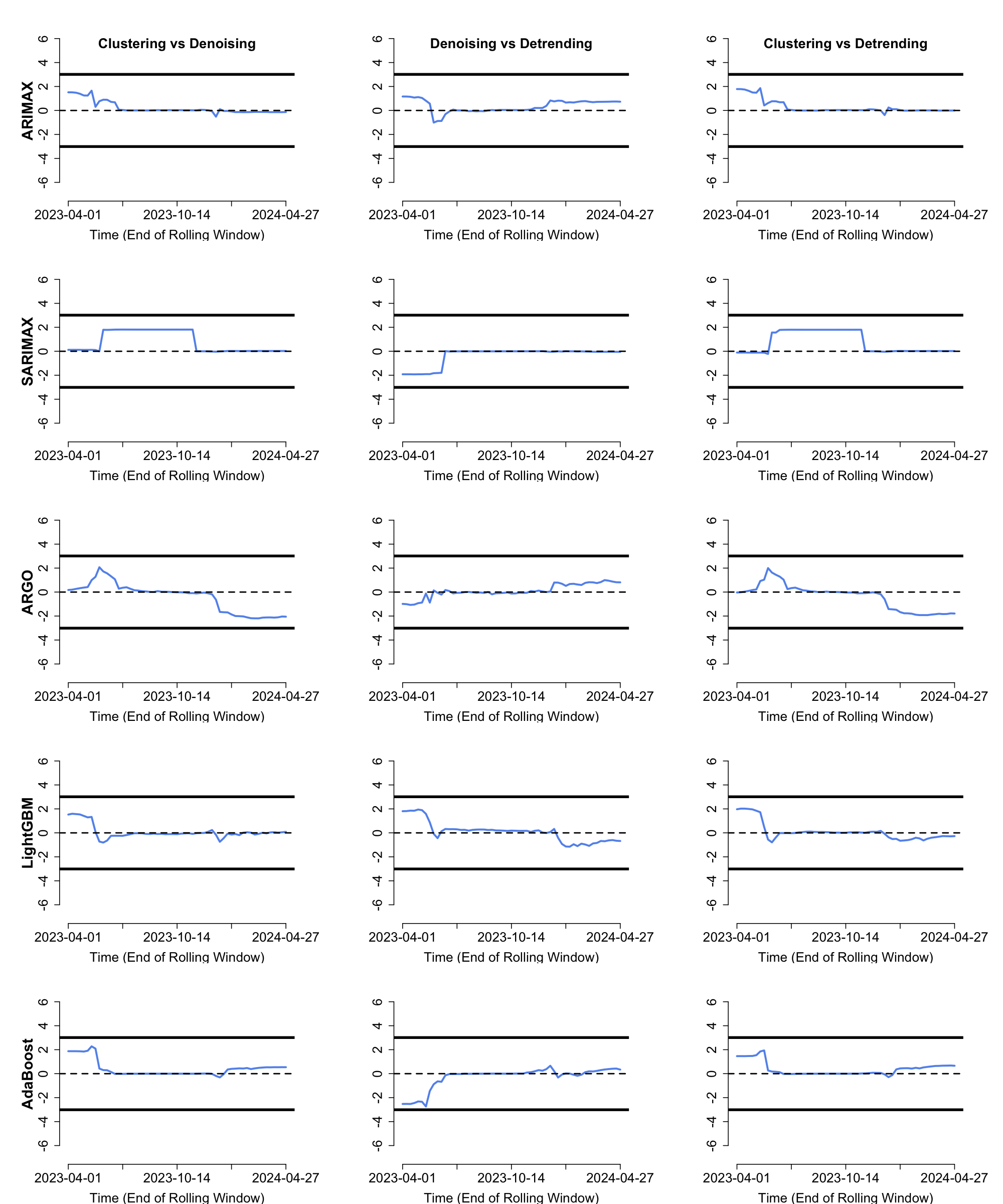}
    \caption{Horizon 0 Fluctuation Test applied to average errors across all locations for each model, using a rolling window of 24 weeks. The blue line shows the path of the test statistic, with critical values indicated by thick black lines. If the path stays within the critical boundaries, we fail to reject the null hypothesis that pairs of preprocessing steps perform equally well at all time points. If the path crosses the critical values, one model outperformed the other at some point in time.}
    \label{fig:fluctuation_h0}
\end{figure}

\begin{figure}[H]
    \centering
    \includegraphics[width=1\linewidth]{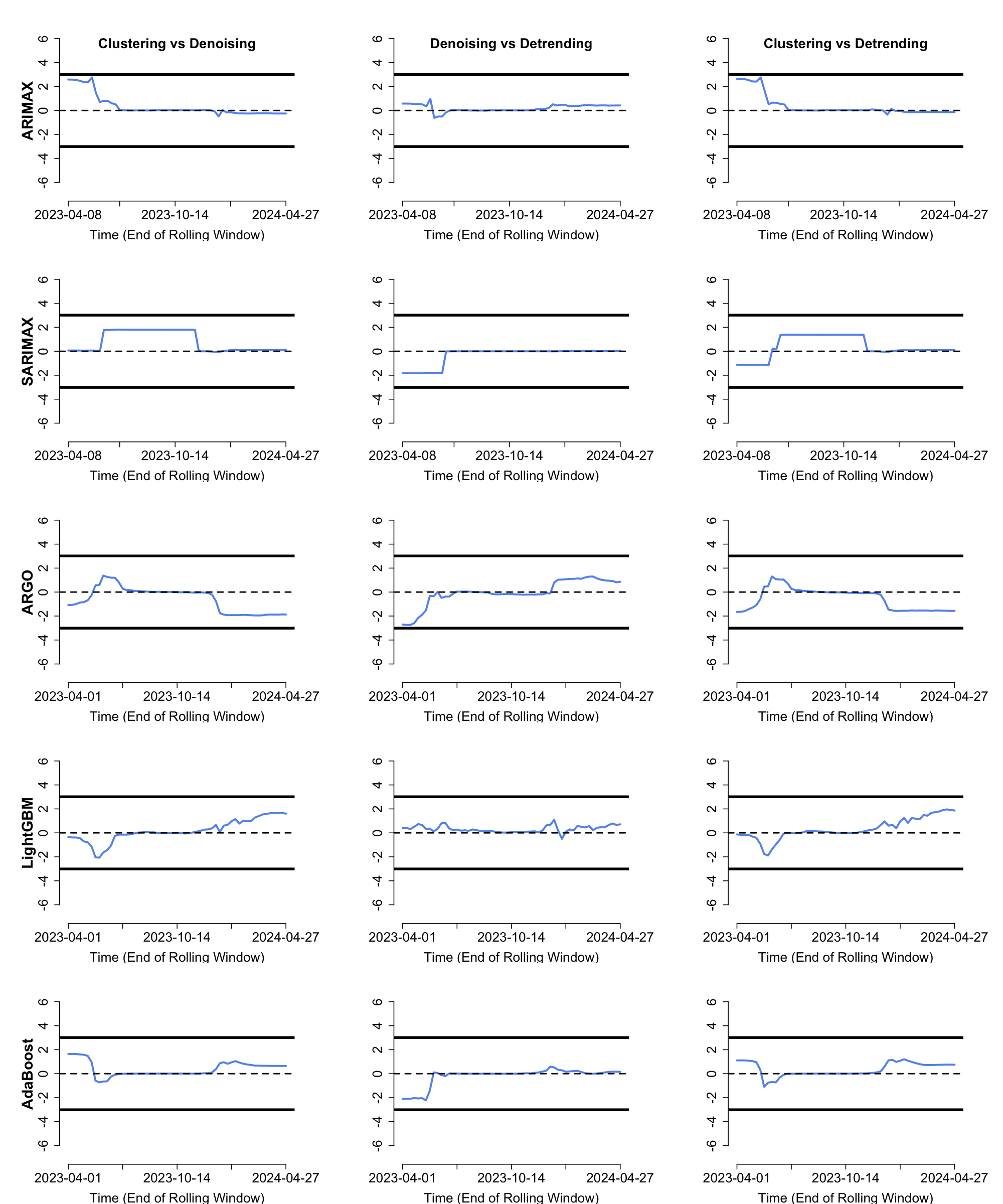}
    \caption{Horizon 1 Fluctuation Test applied to average errors across all locations for each model, using a rolling window of 24 weeks. The blue line shows the path of the test statistic, with critical values indicated by thick black lines. If the path stays within the critical boundaries, we fail to reject the null hypothesis that pairs of preprocessing steps perform equally well at all time points. If the path crosses the critical values, one model outperformed the other at some point in time.}
    \label{fig:fluctuation_h1}
\end{figure}

\begin{figure}[H]
    \centering
    \includegraphics[width=1\linewidth]{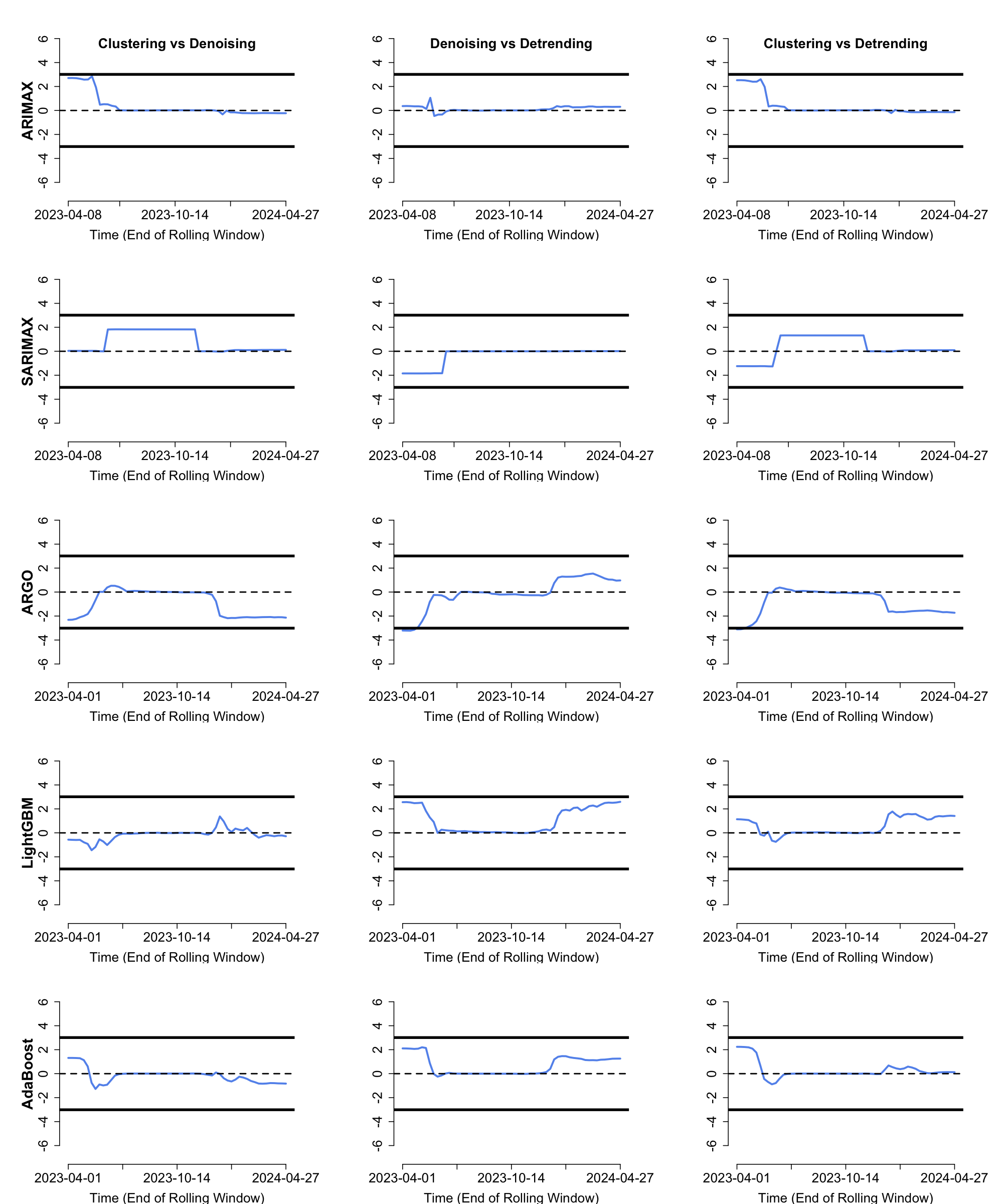}
    \caption{Horizon 2 Fluctuation Test applied to average errors across all locations for each model, using a rolling window of 24 weeks. The blue line shows the path of the test statistic, with critical values indicated by thick black lines. If the path stays within the critical boundaries, we fail to reject the null hypothesis that pairs of preprocessing steps perform equally well at all time points. If the path crosses the critical values, one model outperformed the other at some point in time.}
    \label{fig:fluctuation_h2}
\end{figure}

\begin{figure}[H]
    \centering
    \includegraphics[width=1\linewidth]{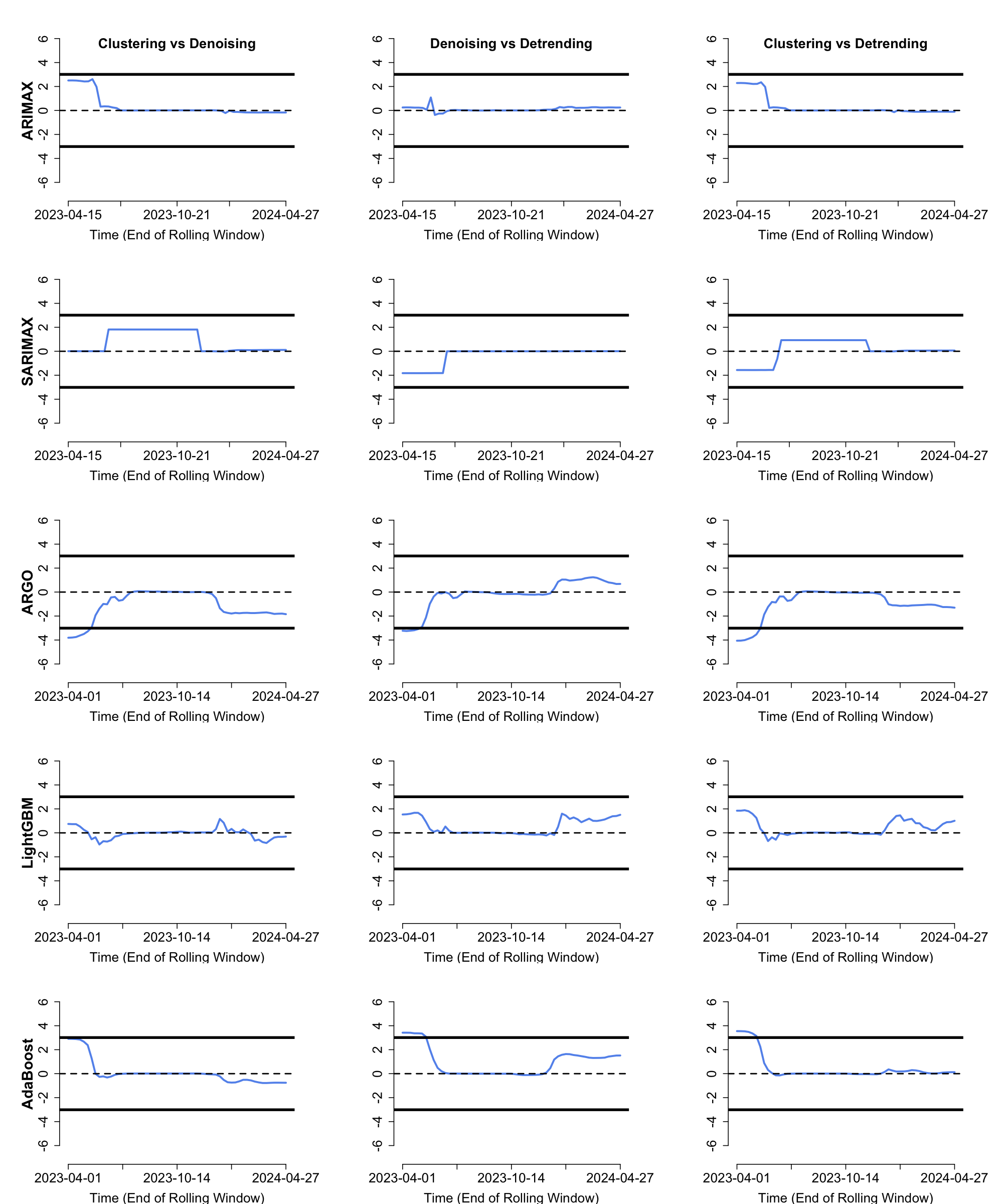}
    \caption{Horizon 3 Fluctuation Test applied to average errors across all locations for each model, using a rolling window of 24 weeks. The blue line shows the path of the test statistic, with critical values indicated by thick black lines. If the path stays within the critical boundaries, we fail to reject the null hypothesis that pairs of preprocessing steps perform equally well at all time points. If the path crosses the critical values, one model outperformed the other at some point in time.}
    \label{fig:fluctuation_h3}
\end{figure}

\section{Model Specifications}

\subsection{Denoising Models}
\label{app:denoising}

For the SSA filter, we specify the number of lagged vectors $L$ to construct from the input time series and the number of leading components $k$ to retain. \citet{hassani2007singular} specify that $2 \leq L \leq T/2$, where $T$ is the training window size of our iterative denoising algorithm, and that $L$ should be proportional to the periodicity of the time series. Since flu-related time series typically exhibit annual seasonality, we set $T=104$ weeks and $L=52$ weeks. SSA decomposes the series into a set of eigentriples representing trend, oscillatory patterns, and noise. We choose $k=3$ leading components, as the first component captures the trend and the next two form a pair that represents the annual cycle.

For the WT filter, we follow the specifications from \cite{fenga2020filtering}: using hard thresholding, symmetric Daubechies wavelet with 8 vanishing moments (SymmletS8), ``chosen for its near symmetry and compact support'', periodic boundaries, 4-level wavelet decomposition. We optimize for the threshold value between 0.1 and 2 based on the paper's reported optimal values.

\subsection{Forecasting Models}
\label{app:model_specs}

Traditional time series models leverage past values and external predictors and are implemented as follows. AutoRegressive Integrated Moving Average with eXogenous variables (ARIMAX) is an extension of the standard ARIMA model that incorporates external predictors. We choose a simple ARIMAX(1,1,1) model with first-order differencing, and a seasonal ARIMAX(1,1,1) model with yearly seasonal differencing. ARGO is implemented with 52 lags of the target to capture annual seasonality. These statistical models are trained on a rolling window of 104 weeks.

Tree-based machine learning approaches flexibly capture nonlinear relationships and interactions and use the following specifications. LightGBM and AdaBoost are trained on an expanding window with a minimum of four years of past data to mimic real-time forecasting and ensure sufficient training data, using all observations prior to each prediction date. For LightGBM, we include up to 52 weeks of target lags to capture seasonal dependencies and up to 4 weeks of exogenous variable lags to model short-term effects. We configure the model with the gradient-based one-side sampling (GOSS) boosting strategy, a maximum tree depth of 5, 300 boosting iterations, and a learning rate of 0.5. For AdaBoost, we restrict the input to 1–4 weeks of target and exogenous lags to emphasize recent dynamics, and use decision trees of maximum depth 3 as base learners with 500 boosting iterations and a learning rate of 0.5. These configurations are designed to balance predictive flexibility with efficiency, limiting overfitting while maintaining computational feasibility in the rolling forecasting setup that requires frequent retraining.

\end{document}